\documentclass[a4paper,11pt]{article}

\pdfoutput=1 

\usepackage{jheppub}

\usepackage{mathrsfs}
\usepackage{comment}
\usepackage{centernot}
\usepackage{mathtools}
\usepackage{stmaryrd}

\usepackage{latexsym,amssymb,amstext,amsmath}
\usepackage{hyperref}
\usepackage{graphbox,graphicx}
\usepackage{graphicx} 
\usepackage{subcaption}
\usepackage{amsthm}
\usepackage{esint}
\usepackage{float}
\usepackage{xcolor}

\usepackage{slashed}

\def\bea {\begin{eqnarray}}
\def\eea {\end{eqnarray}}

\def\beq{\begin{equation}}
\def\eeq{\end{equation}}
\def\beqa{\begin{eqnarray}}
\def\eeqa{\end{eqnarray}}

\theoremstyle{definition}

\numberwithin{equation}{section}

\title{\boldmath Bulk-boundary entanglement correspondence  and the Ryu-Takayanagi conjecture in an $AdS_2/CFT_1$ setup}

\author{Gabriel Lopes Cardoso${}^1$, Bernardo Moniz Martins${}^2$, Suresh Nampuri${}^1$}

\affiliation{${}^1$ Center for Mathematical Analysis, Geometry and Dynamical Systems,\\
Department of Mathematics, Instituto Superior T\'ecnico, Universidade de Lisboa,
1049-001 Lisboa, Portugal}
\affiliation{${}^2$ Department of Physics, Instituto Superior T\'ecnico, Universidade de Lisboa,
1049-001 Lisboa, Portugal}

\emailAdd{gabriel.lopes.cardoso@tecnico.ulisboa.pt}
\emailAdd{bernardo.moniz.martins@tecnico.ulisboa.pt}
\emailAdd{nampuri@gmail.com}

\abstract{Using recent developments in expressing one-loop partition functions in Euclidean $AdS_2$ space-times in terms of character integrals, we relate the one-loop effective action for a free field theory in $AdS_2$ (comprised of a  massless scalar field and a massless Majorana fermion field) to the partition function of the de Alfaro-Fubini-Furlan (DFF) conformal quantum mechanics (CQM) models on the two global $AdS_2$ boundaries. The equal number of bosonic and fermionic degrees in the field theory guarantee that the one-loop calculation is free of all UV divergences except a logarithmic one consistent with the expected entanglement entropy behaviour in a CQM. Via a thermofield double representation, we compute the entanglement entropy between two copies of the $CFT_1$ (CQM), each living near one of the two boundaries of global $AdS_2$, in a state at global time $\tau \rightarrow  - \infty$. This entanglement entropy is expressed in terms of the logarithm of the regularised length of a closed particle trajectory infinitesimally near the rim of the Euclidean $AdS_2$ disc. We view this relation between boundary quantum entanglement and a bulk geometrical quantity as the $AdS_2/CFT_1$ version of the Ryu-Takayanagi conjecture in our setup. The boundary entanglement entropy is equal to 4 times the
thermodynamic entropy read off from 
the regularised one-loop effective action in $AdS_2$. Further, we compute the bulk entanglement entropy associated with black hole horizons in Lorentzian $AdS_2$ and show that it precisely matches the boundary entanglement entropy. 
 
 }

\begin{document} 
\maketitle
\flushbottom

\section{Introduction}

We use recent developments in expressing one-loop partition functions in Euclidean $AdS_2$ space-times in terms of character integrals \cite{Anninos:2020hfj,Sun:2020ame,Grewal:2021bsu} to compute the regularised one-loop effective action\footnote{For a generalization of the construction given in \cite{Anninos:2020hfj,Sun:2020ame}
of one-loop partition functions
on spheres and anti-de Sitter spaces in terms of Harish-Chandra characters to $p$-forms, see \cite{David:2021wrw}.} for a conformal free field theory comprised of a real massless scalar field and a massless Majorana fermion field with 
$c = \bar c = \frac32$. 
The regularised one-loop effective action $ \Gamma^{\rm reg}_{\rm 1-loop} (\bar{\epsilon} )$ for this field theory, when expressed as \cite{Banerjee:2010qc,Banerjee:2011jp}
\bea
\Gamma^{\rm reg}_{\rm 1-loop} (\bar{\epsilon} ) = \beta \Delta E -  \Delta S \;,
\eea
gives 
\bea
\Delta S   = \frac14 \, \log \frac{1}{\bar \epsilon}\;\;\;,\;\;\;
\bar{\epsilon} = \frac{\epsilon}{a} \;,
\eea
where $a$ denotes the $AdS_2$ scale factor and $\epsilon$ 
denotes a UV cutoff. By identifying the bulk IR cutoff to be numerically equal to $a^2/\epsilon$,
we trade $\epsilon$ in the above expression for the 
length $L = 2 \pi a^2/\epsilon$  of a closed circular particle geodesic located near the boundary of the Euclidean $AdS_2$ disc. This allows us to express $\Delta S$, computed via a replica trick in the bulk, in terms of a geometrical scale $L$ as
\bea
\Delta S = \frac14 \, \log \frac{L}{2 \pi a}\;.
\label{dSLsum}
\eea

In this note we relate $\Delta S $ to the entanglement entropy between two copies of a $CFT_1$, each living on one of the two boundaries of global $AdS_2$, in a state $|\tau \rangle$ at global time $\tau \rightarrow - \infty$. The $CFT_1$ is the conformal quantum mechanics model of de Alfaro-Fubini-Furlan (DFF) \cite{deAlfaro:1976vlx} at the specific coupling value $g = 3/4$. It arises by noting that the unregularised one-loop effective action in Euclidean $AdS_2$ can be expressed in terms of the partition function of the DFF model, integrated over Schwinger proper time. We construct a state $|\tau \rangle$ that satisfies a Schr\"odinger type equation \cite{Chamon:2011xk,Jackiw:2012ur},
\bea
\mathbb{S} |\tau \rangle = - i \frac{d}{d \tau} |\tau \rangle \;, 
\eea
where $\mathbb{S} = - \frac12 (L_+ + L_-)$ denotes one of the generators of the underlying $so(2,1)$ Lie algebra.
Using a thermofield double representation \cite{Arzano:2023pnf}, we express the $so(2,1)$ generators in terms of 
operators acting on the direct product of two identical independent Hilbert spaces ${\cal H}_L \otimes {\cal H}_R$. Likewise, we express 
the state $|\tau\rangle$ as a linear combination of states in 
${\cal H}_L \otimes {\cal H}_R$.
Following the operator-state correspondence in $AdS_2/CFT_1$ \cite{Sen:2011cn}, we consider the state $|\tau \rightarrow - \infty \rangle$ located on both boundaries of global $AdS_2$.
Denoting this state by $|\delta \rangle$, we consider the pure state density matrix $\hat{\rho} = |\delta\rangle \langle \delta |$ on ${\cal H}_L \otimes {\cal H}_R$. By tracing over ${\cal H}_L$, we compute the associated von Neumann entropy $S_{\rm vN}$. Our expression for $S_{\rm vN}$ agrees with the expression obtained earlier in \cite{Arzano:2023pnf} in a related context.
We establish that, up to $\epsilon$-independent constants,
\bea
\Delta S = \frac14 \, S_{\rm vN} \;.
\eea
The above equations express the von Neumann
entropy in the boundary CQM thermofield double state $|\delta \rangle$ in terms of the logarithm of the length $L$ of a closed circular geodesic located near the boundary of the Euclidean $AdS_2$ disc. We view this relation as the $AdS_2/CFT_1$ version of the celebrated Ryu-Takayanagi conjecture \cite{Ryu:2006bv,Ryu:2006ef} in our setup.

\section{$AdS_2$: dynamics in various coordinate systems}

\subsection{Coordinate systems \label{sec:coord}}

Here we briefly review various coordinate systems for $AdS_2$ space-time.

We begin by recalling that 
the metric describing the near-horizon region of an extremal (BPS) black hole in four space-time dimensions takes the form of a product geometry, with the line element given by
\cite{Sen:2008vm}
\bea
ds^2_4 = a^2 \left( - (r^2 -1) dt^2 + \frac{dr^2}{r^2 -1} \right) + a^2 \left( d \theta^2 + \sin^2 \theta \, d\phi^2 \right) \;.
\label{rnr1}
\eea
The product geometry is 
$AdS_2 \times S^2$, where the $AdS_2$ part 
\bea
ds^2_2 = a^2 \left( - (r^2 -1) dt^2 + \frac{dr^2}{r^2 -1} \right) 
\label{ads2rt} 
\eea
is referred to as `black hole' in $AdS_2$, since it exhibits two horizons at $r = \pm 1$. 
As shown in \cite{Sen:2008vm}, this $AdS_2$ metric arises by carefully taking the extremal limit of the near-extremal near-horizon geometry of a Reissner-Nordstrom black hole in four dimensions.

Let us restrict attention to the coordinate patch $1 < r < + \infty$. Then, using
\bea
x =  \frac12 \, \log \frac{r+1}{r-1} \;\;\;,\;\;\; dx = \frac{dr}{r^2 -1} \;\;\;,\;\;\; 
0 < x <  + \infty \;,
\label{xrc}
\eea
the two-dimensional metric \eqref{ads2rt} takes the following form in the $(t,x)$ coordinate system,
\bea
ds_2^2 = a^2 \frac{\left( - dt^2 + dx^2 \right)}{\sinh^2 x} \;\;\;,\;\;\; 
0 < x <  + \infty \;.
\label{txb}
\eea
On the other hand, using the coordinate transformation 
\bea
r  = \cosh \rho  \;\;\;,\;\;\; 0 < \rho < + \infty \;\;\;,\;\;\; \frac{dr^2}{r^2 -1} = d \rho^2  \;,
\eea
the metric \eqref{ads2rt} can be brought to the form
\bea
ds_2^2 = a^2 \left( - \sinh^2 \rho \, dt^2 + d\rho^2 \right) \;\;\;,\;\;\;  0 < \rho < + \infty \;.
\label{adssinh}
\eea
The coordinate system $(t,x)$ is related to the coordinate system $(t,\rho)$ by 
\bea
\sinh x = \frac{1}{\sinh \rho} > 0 \;.
\eea

In Poincar\'e coordinates, the $AdS_2$ line element takes the form
\bea
ds^2_2 = a^2 \left( - r^2 dt^2 + \frac{dr^2}{r^2}  \right) \;\;\;,\;\;\; 0 < r < + \infty \;,
\label{poincline}
\eea
while in global coordinates the $AdS_2$ line element takes the form
\bea
ds^2_2 = a^2 \left( - \cosh^2 \hat{\sigma} \, d\tau^2 + d \hat{\sigma}^2  \right) \;\;\;,\;\;\; - \infty < \hat{\sigma} < +\infty \;.
\eea
Using 
\bea
\tan \sigma = \sinh \hat{\sigma} \;\;\;,\;\;\;  - \frac{\pi}{2}  < \sigma < \frac{\pi}{2}\;,
\eea
the latter line element can be brought to the form
\bea
ds_2^2 = \frac{a^2}{\cos^2 \sigma} \left( - d \tau^2 + d \sigma^2 \right) \;\;\;,\;\;\; - \frac{\pi}{2}  < \sigma < \frac{\pi}{2}\;,
\eea
and finally, by performing the shift $\theta \mapsto \theta - \frac{\pi}{2}$, the $AdS_2$ line element 
in global coordinates takes the form
\bea
ds_2^2 = \frac{a^2}{\sin^2 \sigma} \left( - d \tau^2 + d \sigma^2 \right) \;\;\;,\;\;\; - \pi  < \sigma < 0\;.
\label{linegl}
\eea

Defining
\bea
U_{\pm} = \frac12 \left( \tau \pm \sigma \right) \;\;\;,\;\;\; v_{\pm} = \frac12 \left( t \pm x \right) \;,
\eea
the $(t,x)$ coordinate patch \eqref{txb} gets mapped to region I of global $AdS_2$ (see Figure \ref{fig:hor}) by the map \cite{Sen:2011cn}
\bea
\tan U_{\pm} =  - \tanh v_{\pm}
\eea
when restricting $U_{\pm}$ to values in the interval $[- \frac{\pi}{4}, \frac{\pi}{4} ]$.

Next, let us consider the line element of Euclidean $AdS_2$ in various coordinate systems.
Performing the Wick rotation  $y = - i t$
 in \eqref{txb}, we obtain the metric on the hyperbolic space
$\mathbb{H}^2_{\rm BH}$, 
\bea
ds_2^2 = a^2 \frac{\left( dx^2 + dy^2 \right)}{\sinh^2 x} \;\;,\;\;\;
x \in \mathbb{R}^+ \;,
\label{wickxy}
\eea
where  
we take $y$ to be a periodic variable, $0 \leq y < 2 \pi$.

Performing the Wick rotation $\varphi = - i t$ in \eqref{adssinh} results in the metric
\bea
ds_2^2 = a^2 \left(  \sinh^2 \rho \, d \varphi^2 + d\rho^2 \right) \;\;,\;\;\; \rho \in \mathbb{R}^+ \;.
\label{rvades2}
\eea
Taking $\varphi$ to be a periodic variable, i.e. $\varphi \equiv \varphi + 2 \pi$,
and performing the coordinate transformation 
\bea
z = e^{i \varphi} \, \tanh \frac{\rho}{2} \;,
\label{coordisc}
\eea
maps the metric \eqref{rvades2} to the Poincar\'e disc metric (c.f. (82) in \cite{Kitaev:2017hnr}),
\bea
ds_2^2 = 4 a^2 \frac{dz \, d \bar z}{(1- |z|^2 )^2} \;\;\;,\;\;\; |z| < 1 \;.
\label{pdiscm}
\eea

And finally, performing the Wick rotations $t \mapsto - i t$ and $\tau \mapsto - i \tau$ in \eqref{poincline}
and \eqref{linegl}, respectively, gives
\bea
ds^2_2 &=& a^2 \left(  r^2 dt^2 + \frac{dr^2}{r^2}  \right) \;\;\;,\;\;\; t \in \mathbb{R} \;\;\;,\;\;\; r \in \mathbb{R}^+\;, 
\label{pmet}\\
ds_2^2 &=& \frac{a^2}{\sin^2 \sigma} \left( d \tau^2 + d \sigma^2 \right) \;\;\;,\;\;\;
\tau \in \mathbb{R} \;\;\;,\;\;\; - \pi  < \sigma < 0\;.
\label{gmet}
\eea

The Euclidean line elements \eqref{pmet} and \eqref{pdiscm} are related 
by the coordinate transformation \cite{Sen:2008yk},
\bea
\frac{1 - \frac{1}{r}  + i \, t}{1 + \frac{1}{r}  - i \, t } = e^{i \varphi} \,
\tanh \frac{\rho}{2} \;, 
\eea
while the Euclidean line elements \eqref{gmet} and \eqref{rvades2}
are related by the map \cite{Sen:2011cn}
\bea
\tan \left( \frac{\sigma + i \tau}{2} \right) = \tanh \left( \frac12 \left( \log \hat{\rho} + i \varphi \right) \right) \;,
\label{mapstauhrv}
\eea
where
\bea
\hat{\rho} = \tanh \frac{\rho}{2} \;.
\eea
When $\hat \rho = 1$, \eqref{mapstauhrv} maps the circle labeled by $\varphi$ to the pair of lines
$\{-\pi, 0\} \times \mathbb{R}$. Namely \cite{Sen:2011cn}, 
the segment 
$\hat{\rho} = 1, - \frac{\pi}{2} < \varphi < \frac{\pi}{2} $ gets mapped to the boundary $\sigma = 0$, while the 
segment 
$\hat{\rho} = 1,  \frac{\pi}{2} < \varphi < \frac{3\pi}{2} $ gets mapped to the boundary $\sigma = - \pi$, with $\varphi =   \frac{\pi}{2}$ mapping to $\tau \rightarrow + \infty$ on both boundaries, while $\tau \rightarrow -\infty$ at $\sigma =0$ and $\sigma =-\pi$ are respective images of  $\varphi = - \frac{\pi}{2}$ and $\varphi = \frac{3 \pi}{2}$.

\subsection{$so(2,1)$ algebra}

In the following, we will be primarily working with the $(t,x)$  and the $(x,y)$ coordinate systems \eqref{txb} and \eqref{wickxy} respectively.

The Killing vector fields $X$ generating the continuous isometries of 
\eqref{txb} (i.e. $\mathcal{L}_X g = 0$) are given by
\begin{equation}
    \begin{cases}
        X_0 = \partial_t \,,\\
        X_\pm = e^{\pm t} \sinh x \left(\partial_x \pm \coth x \, \partial_t \right)\,.
    \end{cases}
    \label{eq: Killing fields}
\end{equation}
They satisfy the $so(2,1)$ algebra
\begin{equation}
    \left[X_0, X_\pm \right] = \pm X_\pm \,, \quad \left[X_+, X_- \right] = 2 X_0 \,.
    \label{eq: so Algebra}
\end{equation}
This algebra admits the quadratic Casimir operator
\begin{equation}
    Q \equiv X_0^2 + \frac{1}{2} X_+ X_- + \frac{1}{2} X_- X_+ \,.
    \label{Qp}
\end{equation}

The Killing vector fields $L$ generating the continuous isometries of 
\eqref{wickxy} are
\begin{equation}
    \begin{cases}
        L_0 = \partial_y \,, \\
        L_\pm =  e^{\mp iy} \sinh x \left( \partial_x \pm i\coth x\, \partial_y  \right) \,.
    \end{cases}
    \label{eq: L Generators}
\end{equation}
They satisfy the $so(2,1)$ algebra
\begin{equation}
    [L_\pm, L_0] = \pm i L_\pm\,, \qquad [L_+, L_-] = 2 i L_0 \,.
    \label{eq: Lagrangian Generators Commutations}
\end{equation}
The quadratic Casimir operator is \cite{Kitaev:2017hnr} 
\begin{equation}
    Q \equiv - L_0^2 + \frac{1}{2} L_+ L_- + \frac{1}{2} L_- L_+ = \sinh^2 x \left( \partial_x^2 + \partial_y^2 \right) \,.
    \label{Qm}
\end{equation}

\subsection{Point-particle Lagrangian and Hamiltonian}

\subsubsection{Bosonic point-particle}

JT gravity on a disc, with certain boundary conditions imposed, can be described in terms of a charged spinless particle propagating on a disk in the
presence of a constant background magnetic field \cite{Maldacena:2016upp,Kitaev:2018wpr,Yang:2018gdb,Saad:2019lba,Mefford:2020vde}.
The boundary conditions are such that the total boundary length $L$ is kept fixed and
on the boundary the dilaton field $\Phi$ has a prescribed constant value $\Phi_b$. Thus, the disc $D$ may be regarded a disc with wiggly boundary conditions immersed in a Poincar\'e disc \cite{Maldacena:2016upp,Kitaev:2018wpr,Saad:2019lba,Penington:2023dql,Penington:2024sum}. 
Owing to the constraint $R = -2$, the JT action becomes
\bea
- \Phi_b \int_{\partial D} \, dx \, \sqrt{h} \,  \left( K - 1 \right) \;,
\eea
where $h$ in the induced metric on the boundary and $K$ is its extrinsic curvature. Using the Gauss-Bonnet theorem, 
this equals \cite{Yang:2018gdb}
\bea
- \Phi_b \left( 2 \pi \, \chi (D) + A - L \right) \;\;\;,\;\;\;  A = \int_D \, d^2 x \, \sqrt{g} \;\;\;,\;\;\;  
L = \int_{\partial D} \, dx \,  \sqrt{h} \;.
\eea
This can be reinterpreted as the action for a charged spinless particle moving in a Poincar\'e disc in the presence of a magnetic field \cite{Yang:2018gdb}.

In the coordinate system \eqref{rvades2}, using $\chi (D) =a$, this evaluates to \cite{Mefford:2020vde}
\bea
- \Phi_b \, a \, \int_{v_i}^{v_f} dv \left( \cosh \rho (v) \, \dot{\varphi}(v) + \sqrt{\dot{\rho}^2(v) + \sinh^2 \rho(v) \, \dot{\varphi}^2(v)}
\right)  \;,
\label{act}
\eea
where the integration is with respect to boundary time $v$, and
where $(\rho(v_i), \varphi (v_i)) = (\rho(v_f), \varphi (v_f)) $.
As is well known, the action \eqref{act} can be brought to the equivalent form $- \int_{v_i}^{v_f} dv \, {\cal L}$ with 
\bea
{\cal L}=  \Phi_b \, a \,  \cosh \rho (v) \, \dot{\varphi}(v) + \frac12 \left( \frac{a^2}{  e(v)} \,
\left( \dot{\rho}^2(v) + \sinh^2 \rho(v) \, \dot{\varphi}^2(v)
\right) +  e(v) \,  \Phi_b^2   \right)  \,.
\label{act2}
\eea
The equation of motion for the einbein $e$ gives (we assume $\Phi_b \, a >0$)
\bea
e (v) = a \, \frac{\sqrt{\dot{\rho}^2(v) + \sinh^2 \rho(v) \, \dot{\varphi}^2(v)}}{\Phi_b } \;,
\label{eeom}
\eea
which when substituted into \eqref{act2} gives back the action \eqref{act}. 

The Schwarzian limit corresponds to taking both $L$ and $\Phi_b$ large, keeping $ L/\Phi_b $ fixed \cite{Kitaev:2018wpr}.
Here, we will be interested in the extremal limit, i.e. 
we will consider the Lagrangian \eqref{act2}
in the limit
 $\Phi_b \rightarrow 0$ with $L$ fixed\footnote{The Lagrangian \eqref{act3} can also be obtained using the approach of  \cite{Penington:2023dql}, by performing the rescaling $t \mapsto \varphi_b \, t$ in eq. (2.13) of  \cite{Penington:2023dql} and subsequently sending $\varphi_b \rightarrow 0$.}, 
 \bea
{\cal L}=  \frac{a^2}{2} \, \frac{1}{ e(v)} \,
\left( \dot{\rho}^2(v) + \sinh^2 \rho(v) \, \dot{\varphi}^2(v)
\right)     \,.
\label{act3}
\eea
Using reparametrization invariance ($e(v) dv = {\tilde e}({\tilde v}) d {\tilde v}$), we set $e=1$.
 In the coordinate system \eqref{wickxy}, this point-particle Lagrangian reads
\bea
{\cal L} = \frac{a^2}{2 \sinh^2 x}
 \left( \dot{x}^2 +  \dot{y}^2 \right) \;.
\eea
The canonical momenta $p_x = \partial \mathcal{L} / \partial \dot x$ and $p_y = \partial \mathcal{L} / \partial \dot y$ are 
\begin{equation}
 \label{ppxy}
    p_x = \frac{a^2 \, \dot{x}}{\sinh^2 x } \;\;\;,\;\;\;    p_y =  \frac{a^2 \, \dot{y}}{\sinh^2 x} \;,
    \end{equation}
 and the associated Hamiltonian $\mathcal{H} = p_x \dot{x} + p_y \dot{y} - \mathcal{L} $ is
\begin{equation}
    \mathcal{H} = \frac{\sinh^2 x}{2 a^2} \left(p_x^2 + p_y^2 \right) \,.
    \label{eq:CHHE}
\end{equation}
Owing to the equation of motion for the einbein $e$, the classical Hamiltonian \eqref{eq:CHHE} vanishes on-shell.

Upon first quantisation, by promoting the momenta $p_{\mu}$ to operators 
$\hat{p}_{\mu} = - i \partial_{\mu}$,
the constraint $\mathcal{H} =0$ becomes the Klein-Gordon equation ${\hat \Delta} \phi =0$ 
for a real scalar field $\phi$ on the hyperbolic plane $\mathbb{H}^2_{\rm BH}$,
with
\bea
2 \hat{\mathcal{H}}= {\hat \Delta} = - \Box = - \frac{\sinh^2 x}{a^2} \left( \partial_x^2 + \partial_y^2 \right) \;.
\label{delxy}
\eea

\subsubsection{Spinning particle}

Let us now consider the Lagrangian for a massless spinning particle $(x^{\mu}, \psi^{\mu})$ 
\cite{vanHolten:1995qt},
\bea
\mathcal{L} = \frac{1}{2 e} \, g_{\mu \nu} \dot{x}^{\mu} \dot{x}^{\nu} + \frac{i}{2}  g_{\mu \nu} \, \psi^{\mu} D \psi^{\nu} +  \frac{i}{e}  g_{\mu \nu} \, \chi \, \psi^{\mu} \dot{x}^{\nu} \;,
\eea
where $g_{\mu \nu}$ denotes the metric \eqref{wickxy} on the hyperbolic plane $\mathbb{H}^2_{\rm BH}$ and where
\bea
D \psi^{\mu} = \dot{\psi}^{\mu} + \dot{x}^{\lambda} \Gamma^{\mu}{}_{\lambda \nu} \psi^{\nu} \;.
\eea
$e$ and $\chi$ describe the einbein and the gravitino, respectively.  
Varying with respect to these two fields yields the constraint equations
\bea
g_{\mu \nu} \dot{x}^{\mu} \dot{x}^{\nu} = 0 \;\;\;,\;\;\;  g_{\mu \nu} \,\psi^{\mu} \dot{x}^{\nu} = 0 \;.
\eea
Upon first quantisation, the above constraints become the
Dirac equation $\slashed {\cal D} \Psi = 0$ for a Dirac spinor \cite{Brink:1976uf} on the hyperbolic plane $\mathbb{H}^2_{\rm BH}$.

\section{Heat kernel and spectral density}

Now let us discuss the heat kernel of the operators ${\hat \Delta}$ and $\slashed {\cal D}$ in the background \eqref{wickxy}.

We begin by considering the operator ${\hat \Delta}$ given in \eqref{delxy}.
The spectrum of the operator ${\hat \Delta}$, augmented by the presence of a magnetic field, has been discussed in 
different coordinate systems \cite{10.1063/1.526781,10.1063/1.530850,Pioline:2005pf,Banerjee:2010qc,Kitaev:2017hnr,Kitaev:2018wpr,
Mefford:2020vde}. In the absence of a magnetic field, the eigenfunctions $f$ of \eqref{delxy},
\bea
{\hat \Delta} f (x,y) = E \, f(x,y) \;,
\label{eigf}
\eea
are $\delta$-normalizable, with eigenvalues $E$ given by $a^2 \, E = j (1-j)$, with $j = \frac12 + i \nu, \, \nu \in \mathbb{R}^+$ \cite{Kitaev:2017hnr}. The eigenfunctions are given by 
\bea
f(x,y) &=& \, e^{- i k y} \, \phi(\nu, k, x) \;\;\;,\;\;\; k \in \mathbb{Z} \;, \nonumber\\
\phi(\nu, k, x) &=& 
N_{k,\nu} \, e^{- |k| x}  \, \left( 1 - e^{-2x} \right)^{\frac12 + i \nu} \, F\left( |k| + \frac12 + i \nu , \frac12 + i \nu, 1 + |k|, e^{-2x} \right) \;,
\eea
where $F$ denotes the regularised hypergeometric function, 
\bea
F(a, b; c; z) = \frac{1}{\Gamma (c)} \, \text{}_2 F_1 (a,b;c;z) \;,
\eea
and $N_{k,\nu}$ denotes a normalization constant.
Performing the coordinate change  $z = e^{-2x}$, $\phi$ becomes
\bea
\phi(\nu, k, z) =
N_{k,\nu} \, z^{|k|/2}  \, \left( 1 - z \right)^{\frac12 + i \nu} \, F\left( |k| + \frac12 + i \nu , \frac12 + i \nu, 1 + |k|, z \right) \;,
\eea
and by requiring its $\delta$-normalizability, i.e. 
\begin{equation}
  2 \pi \,    \int_0^1 \frac{2 dz}{(1-z)^2} \phi^*(\nu, k, z) \, \phi(\nu', k', z) dz  = \delta(\nu-\nu')  \,,
    \label{eq: Normalization Integral}
\end{equation}
the normalization constant $N_{k,\nu}$ is determined to be \cite{Kitaev:2017hnr}
\bea
|N_{k,\nu}|^2 = \frac{1}{8 \pi^2}  \left| \frac{\Gamma\left(\frac{1}{2} -k -i\nu \right)\Gamma\left(\frac{1}{2}  -i\nu \right)}{\Gamma(-2i\nu)}  \right|^2 \;,
\eea
which can also be written as \cite{Banerjee:2010qc}
\bea
|N_{k,\nu}|^2 = \frac{1}{2\pi}  \left| \frac{\Gamma\left(\frac{1}{2} -k -i\nu \right)}{\Gamma(i\nu)}  \right|^2  \;.
\eea
At $k=0$ this evaluates to
\bea
|N_{0,\nu}|^2 =  \frac{ \nu \, \tanh ( \pi \nu ) }{2 \pi} \;.
\eea

The eigenvalue $E$ equals $E = \frac{1}{a^2} \left( \frac14 + \nu^2 \right), \, \nu \in \mathbb{R}^+$.
Introducing the heat kernel $K_B(s)$ of the operator $\hat \Delta$,
\bea
K_B(s) = \int_{\frac{1}{4 a^2}}^{+ \infty} \, dE \, e^{- E s} \, \hat{\rho} (E) \;,
\eea
where $\hat{\rho} (E)$ denotes the spectral density, we obtain the relation
\bea
K_B(s) = \frac{1}{a^2} \int_{0}^{+ \infty} \, d\nu \, e^{- (\frac14 + \nu^2)  {\bar s}} \, \varrho (\nu) \;,\;
\varrho (\nu) = 2 \nu \, \hat{\rho} \left(E =  \frac{1}{4a^2}  + \frac{\nu^2}{a^2} \right)\;,\;
{\bar s} = \frac{s}{a^2} \;.
\eea
The spectral density $\varrho(\nu)$ equals \cite{Kitaev:2018wpr}
\bea
\varrho (\nu) = V \, | \phi (\nu, k=0, x=0) |^2 = V \, |N_{0,\nu}|^2 = \frac{V}{2 \pi}  \, \nu \, \tanh ( \pi \nu ) \;.
\label{exprho}
\eea
Stripping off the factor $V/2 \pi$, we define
\bea
\rho_B(\nu) = \nu \, \tanh ( \pi \nu ) =  \nu \, \frac{\sinh 2\pi \nu }{\cosh 2 \pi \nu + 1 } \;,
\label{densbos}
\eea
which satisfies $\rho_B (- \nu) = \rho_B (\nu)$. $V$ denotes the regularised volume computed from \eqref{wickxy} in the presence of a cutoff 
$0< x_0 \ll 1$, 
\bea
V = \int_0^{2 \pi} \left(  \int_{x_0}^{+ \infty}  \sqrt{g} \,  dx \right) dy = 2 \pi  a^2 \left( \coth x_0 -1 \right) \;\;\;,\;\;\; 0 < x_0 \ll 1 \;.
\label{Vreg}
\eea
Thus, we obtain for the heat kernel  of the operator $\hat \Delta$, 
\bea
K_B(s) = \frac{V}{2 \pi a^2} \int_{0}^{+ \infty} \, d\nu \, e^{- (\frac14 + \nu^2)  {\bar s}} \, \rho_B(\nu) \;\;\;,\;\;\; \rho_B(\nu) = \nu \, \tanh ( \pi \nu ) 
\;\;\;,\;\;\;
{\bar s} = \frac{s}{a^2} \;.
\label{Kb}
\eea
in agreement with \cite{Pioline:2005pf,Banerjee:2010qc}.

Next, let us consider the Dirac operator $\slashed {\cal D}$ in the background \eqref{wickxy}.
Its spectrum
has been discussed in 
various coordinate systems \cite{10.1063/1.526781,Camporesi:1995fb,Pioline:2005pf,Banerjee:2010qc,Kitaev:2017hnr,Kitaev:2018wpr,
Mefford:2020vde}.
The associated spectral density $\rho_F(\nu)$ 
is given by \cite{Pioline:2005pf,Banerjee:2010qc}
\bea
\rho_F (\nu) 
 = 2  \nu \, \coth ( \pi \nu ) =  2  \nu \, \frac{\sinh 2\pi \nu }{\cosh 2 \pi \nu -1} \;,
 \label{densrhof}
\eea
which satisfies $\rho_F (- \nu) = \rho_F (\nu)$. 
The heat kernel of the Dirac operator in the background \eqref{wickxy} reads \cite{Banerjee:2010qc}
\bea
K_F(s) = -  \frac{V}{2 \pi a^2} \int_{0}^{+ \infty} \, d\nu \, e^{- \nu^2  {\bar s}} \, \rho_F(\nu) \;\;\;,\;\;\; \rho_F(\nu) = 2  \nu \, \coth ( \pi \nu ) 
\;\;\;,\;\;\;
{\bar s} = \frac{s}{a^2} \;.
\label{Kf}
\eea

In order to compute the one-loop effective action, we will work \cite{Anninos:2020hfj,Sun:2020ame,Grewal:2021bsu} with the 
Fourier integral $W_{B,F}(u)$ of the spectral density  $\rho_{B,F}(\nu)$, 
\bea
W_{B,F}(u) \equiv  \int_{-\infty}^{+\infty}
d\nu \, \rho_{B,F}(\nu) \, e^{i \nu u } \;.
\label{wu}
\eea
We note that formally, $W_{B,F}(u)$ has the property $W_{B,F} (-u) = W_{B,F} (u)$ due to $\rho_{B,F}(-\nu) = \rho_{B,F}(\nu)$. 
However, the expression \eqref{wu} is ill-defined: 
since the spectral densities \eqref{exprho} and \eqref{densrhof} do not decay to zero as $\nu \rightarrow \pm \infty$, the Fourier integral \eqref{wu} is divergent. This can be dealt with \cite{Anninos:2020hfj,Sun:2020ame} by suitably deforming the integration contour in \eqref{wu} to run along the positive imaginary axis, 
as we will review in Appendix \ref{sec:rewreff}.
Then,
for the spectral density \eqref{densbos}, 
$W_B(u)$ evaluates to 
\bea
W_B (u)= - e^{- \frac12 u}  \,
\frac{1 + e^{-u}}{(1-e^{-u})^2}  \;,
\label{Wbos}
\eea
which satisfies $W_B (-u) = W_B (u)$. For the spectral density \eqref{densrhof}, 
$W_F(u)$ evaluates to 
\bea
W_F(u) =  - 4   \,
\frac{e^{-u}}{(1-e^{-u})^2}  \;,
\label{Wferm}
\eea
which  satisfies $W_F (-u) = W_F (u)$.

Note that both $W_B$ and $W_F$ have a double pole at $u=0$, 
\bea
W_B(u) &=& - \frac{2}{u^2} - \frac{1}{12} + {\cal O}(u^2)  \;, \nonumber\\
W_F(u) &=& - \frac{4}{u^2} + \frac{1}{3} + {\cal O}(u^2)  \;.
\label{dpW}
\eea

\section{One-loop effective action}

As shown in \cite{Anninos:2020hfj,Sun:2020ame,Grewal:2021bsu}
the one-loop effective action can be expressed in terms of the Fourier integral $W(u)$ of 
the spectral density $\rho (\nu)$. 

The unregularised one-loop effective action suffers from both infrared and ultraviolet divergences, and hence needs to be regularised. The resulting regularised one-loop effective action, which was obtained in \cite{Anninos:2020hfj,Sun:2020ame,Grewal:2021bsu},
will be reviewed in Appendix \ref{sec:regonel}.
It takes the form 
\bea
\Gamma^{\rm reg}_{B, \rm 1-loop} = -  \frac{V}{8 \pi a^2}
 \int_{\mathbb{R} + i \delta} 
\frac{du}{\sqrt{\bar{\epsilon}^2 + u^2}} \, W_B (u) \, e^{- \frac12 \sqrt{\bar{\epsilon}^2 + u^2}} \, e^{- \kappa u} 
\label{gamWu3b}
\eea
for a real Klein-Gordon field, and 
\bea
\Gamma^{\rm reg}_{F, \rm 1-loop} =   \frac{V}{8 \pi a^2}
 \int_{\mathbb{R} + i \delta} 
\frac{du}{\sqrt{\bar{\epsilon}^2 + u^2}} \, W_F (u)  \, e^{- \kappa u} 
\label{gamWu3f}
\eea
 for a complex Dirac field. Here, $\bar \epsilon = \epsilon/a$, where $\epsilon$ denotes the UV regulator, while $\kappa$ denotes the IR regulator.

Let us consider the combination  
\bea
\Gamma^{\rm reg}_{B, \rm 1-loop} +  \frac12 \, \Gamma^{\rm reg}_{F, \rm 1-loop} = 
-  \frac{V}{8 \pi a^2}
 \int_{\mathbb{R} + i \delta} 
\frac{du}{\sqrt{\bar{\epsilon}^2 + u^2}} \left(  W_B (u) \, e^{- \frac12 \sqrt{\bar{\epsilon}^2 + u^2}} 
- \frac12 \, W_F (u)  \right)  e^{- \kappa u} \;. \nonumber\\
\label{regcomb}
\eea
If, for the time being, we proceed formally by switching off both the UV and the IR regulator, we obtain the following unregularised expression for this combination, 
\bea
\Gamma^{\rm unreg}_{B, \rm 1-loop} + \frac12 \Gamma^{\rm unreg}_{F, \rm 1-loop} &=& 
-  \frac{V}{4 \pi a^2}
 \int_0^{+\infty}
\frac{du}{u} \left(  W_B (u) \, e^{- \frac12 u }
- \frac12 \, W_F (u)  \right)  \nonumber\\
&=& -  \frac{V}{4 \pi a^2}
 \int_0^{+\infty}
\frac{du}{u} \left(   \frac{e^{-u}}{1-e^{-u}}   \right) .
\label{unregeffdvv}
\eea
We note that this is the combination for which the double pole in \eqref{dpW} precisely cancels out.
By taking this combination, we halved the contribution coming from the Dirac field. Thus, the on-shell degrees of freedom are constituted of a massless scalar field and a massless Majorana fermion. We refer to \cite{GonzalezLezcano:2023uar} for a recent proposal on the construction of a supersymmetric basis for scalar and spinor fields in  an Euclidean $AdS_2$ space-time.

In the following, let us first relate the unregularised one-loop effective action \eqref{unregeffdvv}
to the partition function of a conformal quantum mechanics model. Subsequently, we will discuss the 
regularised one-loop effective action.

\subsection{Conformal quantum mechanics from unregularised one-loop effective action \label{sec:cqmun1}}

The left hand side of \eqref{unregeffdvv} denotes an unregularised one-loop effective action, which we denote by
$-\log Z^{\rm unreg}_{\rm 1-loop}$. We then write \eqref{unregeffdvv} as 
\bea
\log Z^{\rm unreg}_{\rm 1-loop}  = \frac{V}{2 \pi a^2} \,
\int_0^{+\infty} \frac{du}{2u} \, 
Z_{\rm CQM}(u) \;, 
\label{HZun}
\eea
where
\bea
Z_{\rm CQM}(u) =   \frac{e^{-u}}{1-e^{-u}} =  \sum_{n=0}^{\infty} e^{- (n +1) u}  \;\;\;,\;\;\; u > 0 \;.
\label{zcqmdff}
\eea
We now regard $Z_{\rm CQM}(u)$ as computing the trace over eigenstates of a quantum mechanics model, 
\bea
Z_{\rm CQM}(u) = \text{Tr} \, q^{\hat N + 1} \;\;\; ,\;\;\; {\hat N} = a^{\dagger } a \;\;\;,\;\;\; {\hat N} |n \rangle = n |n \rangle  \;\;\; n \in \mathbb{N}_0 \;\;\;,\;\;\; q = e^{-u} \;.
\label{dffmod}
\eea
Then, formally we obtain
\bea
\log Z^{\rm unreg}_{\rm 1-loop} =   \frac{V}{2 \pi a^2} \,\int_0^{+\infty} \frac{du}{2u} \,  \text{Tr} \, q^{R}
\;\;\;,\;\;\; R 
= {\hat N} + 1  \;.
\label{clopenZ}
\eea
On the LHS, we have the logarithm of a one-loop partition function in an $AdS_2$ background, 
 while 
the RHS is read as the partition function of a quantum mechanics model (CQM), which is being integrated over $u$. Formally, the RHS is, up to a sign and up to the  factor $\frac{V}{2 \pi a^2}$, 
the unregularized 1-loop effective action  of the CQM (cf. eq. (1.18) in \cite{Vassilevich:2003xt}).
Thus, formally, we may view \eqref{HZun} as a correspondence between two 1-loop effective actions, a gravitational one on the LHS and a CQM one on the RHS.

Below we will show that the  quantum mechanics model \eqref{dffmod}  can be identified with the conformal quantum mechanics model of de Alfaro-Fubini-Furlan (DFF) at a specific value of its coupling constant \cite{deAlfaro:1976vlx}.

Below we will make use of the constant term in the Laurent expansion of $H(u) \equiv  - Z_{\rm CQM}(u) $ around $u=0$,
\bea
\frac12 \, H(u) = \sum_{n=0}^{\infty} \, b_n \, u^{n-2} \;\;\;,\;\;\; 
b_0 = 0 \;\;\;,\;\;\; b_1   =  - \frac12 \;\;\;,\;\;\; b_2 = \frac14 \;.
\label{laurH}
\eea

\subsection{Regularised one-loop effective action}

As we will review in Appendix \ref{sec:evreg1}, 
the regularised one-loop effective action \eqref{regcomb}, which we denote by 
$\Gamma^{\rm reg}_{\rm 1-loop} (\bar{\epsilon} )$, can be computed in terms of the 
character zeta function 
\bea
\zeta (z) \equiv \frac{1}{\Gamma(z)} \int_0^{+ \infty} \frac{du}{u} \, u^z \, H (u) \;,
\eea
and reads
\bea
\Gamma^{\rm reg}_{\rm 1-loop} (\bar{\epsilon} ) = 
-\log Z^{\rm reg}_{\rm 1-loop} (\bar{\epsilon} ) = \frac{V}{2 \pi a^2} \left[ - b_2 \log \left( \frac{e^{\gamma} \, \bar{\epsilon}}{2} \right) 
+ \frac12 \zeta' (0) - \frac18 
\right] \;,\; \bar{\epsilon} = \frac{\epsilon}{a} \;, 
\label{regpartit}
\eea
where $b_2$ denotes the constant term in the Laurent expansion \eqref{laurH} and $\gamma$
denotes the Euler-Mascheroni constant.
We recall that $V$ denotes the regularised volume of $\mathbb{H}^2_{\rm BH}$ given by \eqref{Vreg}, which depends on the regulator $x_0$. Thus, $\Gamma^{\rm reg}_{\rm 1-loop} (\bar{\epsilon} ) $ depends on two regulators, namely on $x_0$ as well as 
on the UV regulator $\epsilon$, but it does not depend on the IR regulator $\kappa$ \cite{Sun:2020ame}. The value $\zeta' (0)$ is finite and given in Appendix \ref{sec:evreg1}. We note that the dependence on $\bar \epsilon$ is solely contained in the log-term, i.e. there are no poles in $\bar \epsilon$. This is a consequence of the particular field configuration that we chose, namely one massless scalar field and one massless Majorana fermion. Each of them contributes a term $1/{\bar \epsilon}^2$ to the regularised one-loop effective action, but with an opposite sign, so that these poles cancel out in the regularised one-loop effective action, leaving only the logarithmic divergent term.

Following \cite{Banerjee:2010qc,Banerjee:2011jp}, we write the regularised one-loop effective action \eqref{regcomb} as
\bea
\Gamma^{\rm reg}_{\rm 1-loop} (\bar{\epsilon} ) = V \,  \Delta L_{\rm eff} \;.
\label{VLreg}
\eea
The dependence on the regulator $x_0$ is contained in $V$ only.
The $x_0$ independent term in this expression reads
\bea
- 2 \pi  a^2 \,  \Delta L_{\rm eff} =  b_2 \log \left( \frac{e^{\gamma} \, \epsilon}{2 a} \right) 
- \frac12 \zeta' (0) + \frac18  \;.
\label{x0indt}
\eea
As in \cite{Banerjee:2010qc,Banerjee:2011jp}, we
interpret the regularised one-loop partition function $\Gamma^{\rm reg}_{\rm 1-loop} (\bar{\epsilon} )$
as (see eq. (3.1) in \cite{Banerjee:2011jp})
\bea
\Gamma^{\rm reg}_{\rm 1-loop} (\bar{\epsilon} ) = \beta \Delta E -  \Delta S \;,
\label{freethermo}
\eea
where the term proportional to $\coth x_0$ is identified with $\beta \Delta E$, while the 
$x_0$ independent term is identified with $\Delta S$. Here, $\beta = 2 \pi a /\sinh x_0$ is the inverse temperature given by the length $L$ of the
boundary of $AdS_2$ parametrized by $y$ (cf. \eqref{wickxy} and \eqref{x0} below). Using
$V = \beta \, a \, \cosh x_0 - 2 \pi a^2$, we infer 
\bea
\Delta E &=& a \, \cosh x_0 \, \Delta L_{\rm eff} \;, \nonumber\\
\Delta S &=& b_2  \log \frac{a}{\epsilon} = \frac14 \, \log \frac{a}{\epsilon} \;,
\eea
where we have dropped an additive constant in $\Delta S$. 

Writing $\Delta S$  as 
\bea
\Delta S = 
\frac{c}{6} \, \log \frac{a}{\epsilon}  \;\;\;,\;\;\; c =  1 + \frac12 = \frac32\,,
\label{dels}
\eea
suggests an interpretation of $\Delta S$ as a vacuum entanglement entropy
in an Euclidean CFT  with central charge $c = \bar c$ living on $\mathbb{R}^2$,
as follows.

\begin{figure}
\centering
\includegraphics[scale=.4]{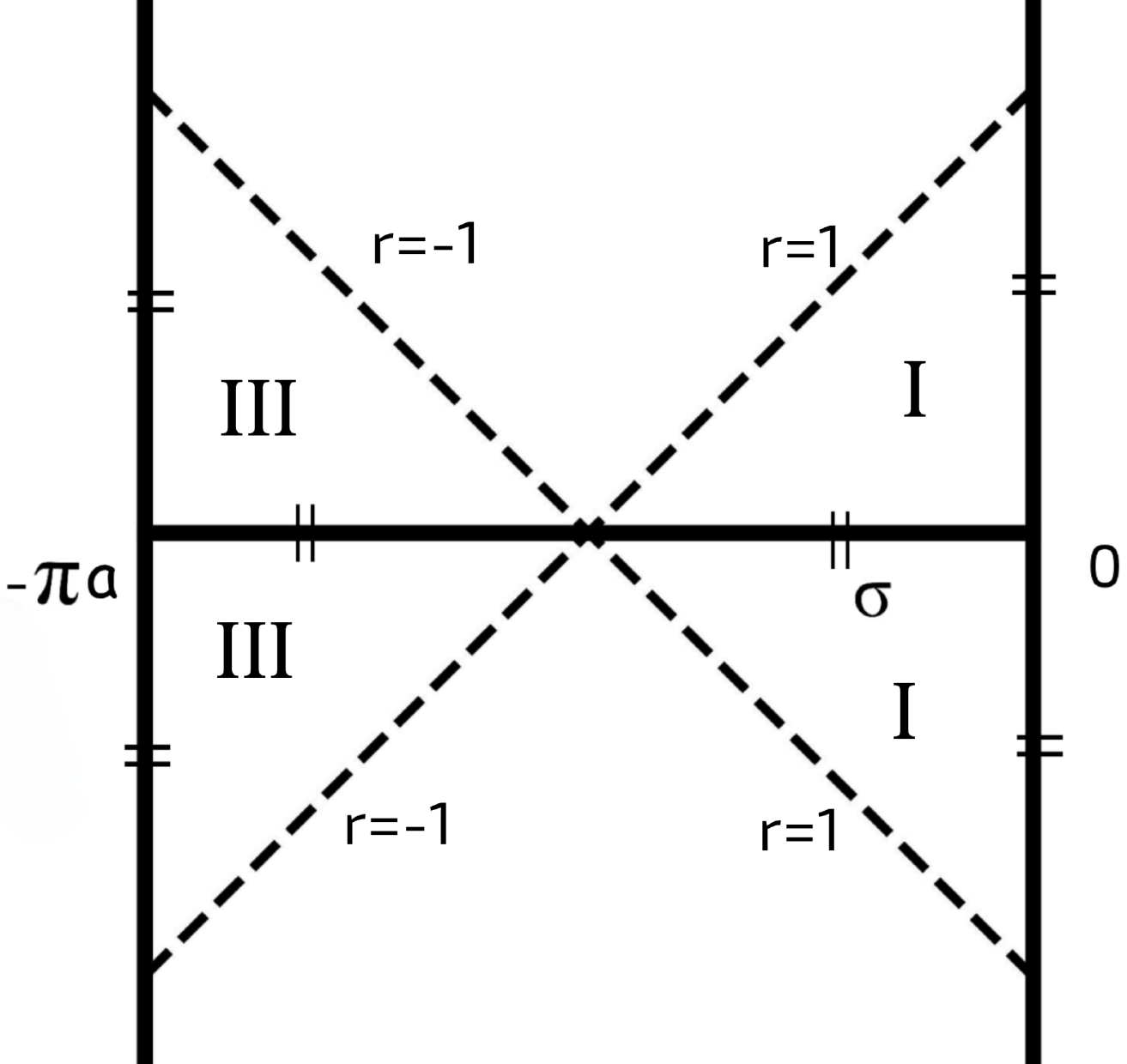}
\caption{\small\textit{Location of the black hole horizons in global $AdS_2$.}}
  \label{fig:hor}
\end{figure}

The black hole horizons in the metric \eqref{rnr1} are null surfaces corresponding to $r=1$ and $r=-1$. They lie along the diagonals in the global $AdS_2$ strip of width $\pi a$ and separate regions $I$ and $III$ from global future and past time-like infinities at $\tau \rightarrow \pm \infty$ \cite{Sen:2008vm}, as depicted in Figure \ref{fig:hor}. 
We divide each region into two equal isosceles right angled triangles by a horizontal line through the diagonal intersection. Each of these triangles has a hypotenuse along a null surface, while the perpendicular to the base lies along an edge of the strip and is of length $\pi a$. 
Each of the 2 diagonals is composed of 2 equal length hypotenuses, one corresponding to  $r=1$ and the other to an $r=-1$ null surface. One of these hypotenuses is porous towards the flow of information towards future timelike infinity, while the other allows information to flow in from past timelike infinity. Subdividing each diagonal into 2 hypotenuses and tracing out the field theory states living  on one of them results in an entanglement entropy associated with the CFT living on the remaining hypotenuse. We can thus view the entanglement entropy across each hypotenuse as that of a CFT subsystem of length $\ell= \frac{1}{\sqrt{2}} \pi a$ nestled within a system of size ${\hat L}=\sqrt{2} \pi a$ with central charge $c=\bar{c}$  at zero temperature. Following  eq. (23) in
\cite{Calabrese:2004eu}, we have, upto additive constants, 
\bea
{\cal S} =  \, \frac{c}{6} \log \left(  \frac{\hat{L}}{\pi \epsilon} \, \sin \left( \frac{\pi \ell}{\hat L} \right) \right) =
\frac{c}{6} \log \frac{a}{\epsilon}\,=\frac{1}{4} \log \frac{a}{\epsilon}.
\label{entflat}
\eea
Here $\epsilon$ is the short-distance cutoff associated with the diagonal which we will identify with the UV scale in the one-loop effective action. 
This calculation depends purely on the geometrical measures of the triangles and is hence the same for the remaining three isosceles right angled triangles. The total entanglement entropy for the given fields associated with the 4 horizon null surfaces is therefore 
\bea \label{finalent}{\cal S}_{\rm total}= \log \frac{a}{\epsilon}.  \eea
The expression \eqref{entflat} agrees with \eqref{dels}.

The length $L$ of a closed circular particle geodesic located at $0 < x_0 \ll 1$, near the $AdS_2$ boundary $\partial M$, is computed to be
\bea
L = \int_{\partial M } dv \,\sqrt{g_{yy} \, \dot{y}^2} = \frac{2 \pi \, a }{\sinh x_0} \;.
\label{x0}
\eea
We now take the value of the bulk IR regulator $\sinh x_0$ to be numerically equal to the regulator $\epsilon$,
\bea
 \epsilon = a \, \sinh x_0 \;.
 \label{eax0}
 \eea
We obtain
 \bea
 L = \frac{2 \pi a^2}{\epsilon} \;,
 \label{Leps}
 \eea
 enabling $\Delta S$ to be written as 
\bea
\Delta S = \frac14 \, \log \frac{L}{2 \pi a}\;,
\label{dSL}
\eea
which expresses $\Delta S$ in terms of the length $L$ of the closed circular geodesic located at  $0 < x_0 \ll 1$.

Global Euclidean $AdS_2$ can be represented as a strip with two boundaries located at $\sigma =- \pi, 0$, cf. \eqref{gmet}. Using the relation \eqref{HZun}, we assign to each of these boundaries a copy of a conformal 
quantum mechanics model which in the next section we will identify with the DFF model  \cite{deAlfaro:1976vlx} at a specific value of its coupling. 
We will then use the results of \cite{Chamon:2011xk,Jackiw:2012ur,Arzano:2023pnf} to
show that ${\cal S}_{\rm total}=4 \,\Delta S$ can be equated to an entanglement entropy 
between these two copies which constitute a thermofield double description of the DFF model,
computed in a state denoted by $| \delta \rangle$.

\section{The DFF model}

In this section, we review various properties of the spectrum of 
the conformal quantum mechanics model of DFF \cite{deAlfaro:1976vlx}. 
The DFF Hamiltonian $H$ reads
\bea
H = \frac12 \left( p^2 + \frac{g}{q^2} \right) \;\;\;,\;\;\; g > 0 \;,
\eea
where $g$ denotes the dimensionless coupling of the model.
We follow 
\cite{Chamon:2011xk,Jackiw:2012ur}, who have studied this model in great detail.

We consider the $so(2,1)$ algebra \eqref{eq: Lagrangian Generators Commutations}.
We redefine $L_- \rightarrow -L_-$ and define $R = i L_0$,
in which case 
\begin{equation}
    [R, L_\pm] = \pm  L_\pm\,, \qquad [L_-, L_+] = 2 R \,.
    \label{daffalg}
\end{equation}
The generators $R, L_{\pm}$ are related to the generators $H, K, D$ (where $H$ denotes the Hamiltonian, $K$ the conformal boost generator and $D$ the dilation generator) by
\bea
R = \frac12 \left( \frac{K}{a} + a \, H \right) \;\;\;,\;\;\; L_{\pm } = \frac12 \left( \frac{K}{a} - a \, H \right) \pm i \, D \;,
\eea
where $a$ denotes the $AdS_2$ scale (of length dimension one) in \eqref{wickxy}.
The generator $R$ generates a compact subgroup $SO(2)$, and hence its spectrum is discrete,
\bea
R |n \rangle = \left( n + r_0 \right) \, |n \rangle \;\;\;,\;\;\; n \in \mathbb{N}_0 \;\;\;,\;\;\; \langle  m | n \rangle  = \delta_{m,n} \;.
\label{Rgen}
\eea
The state $|0\rangle$ is the $R$-vacuum. 
We set $r_0 = 1$ in view of \eqref{dffmod}. In the DFF model, $r_0$ is expressed in terms of the coupling constant $g$ of the model as
\bea
r_0 = \frac12 \left( 1 + \sqrt{g + \frac14} \right) \;.
\eea
It follows that $g = \frac34$.

The operators $L_{\pm}$ act as ladder operators on $|n\rangle$,
\bea
L_{\pm} | n \rangle = \sqrt{(n+1)(n+1 \pm 1) } \, |n \pm 1 \rangle \;,
\eea
which implies
\bea
|n \rangle = \frac{1}{\sqrt{n! (n+1)!}} \, (L_+)^n  \, |0 \rangle \;.
\eea
The Casimir operator  \eqref{Qm} (where we replaced $L_-$ by $-L_-$)
\bea
Q = R^2 - R -  L_+ L_- \;,
\eea
annihilates the states $| n\rangle$,
\bea
Q |n\rangle = 0 \;.
\eea

We define the operator $\mathbb{S}$ by 
\bea
\mathbb{S} =  \frac12 \left( a \, H  - \frac{K}{a}  \right) \;.
\eea
In the following, we will discuss Schr\"odinger type equations for the operators, $H, R, \mathbb{S}$, each
written with respect to a different time variable.

\subsection{$H, R$ and $\mathbb{S}$ }

Following \cite{Chamon:2011xk,Jackiw:2012ur}, we consider Schr\"odinger type equations for the operators, $H, R, \mathbb{S}$, 
\bea
H |t\rangle = - i \frac{d}{dt} |t\rangle \;, \nonumber\\
R |\varphi \rangle = - i \frac{d}{d {\varphi}} |\varphi \rangle \;,  \nonumber\\
\mathbb{S} |\tau \rangle = - i \frac{d}{d {\tau}} |\tau \rangle \;, 
\label{schrHRS}
\eea
where the time variables $t, \varphi, \tau$ refer to Poincar\'e patch time, black hole time and global time, respectively. 
The associated Euclidean line elements are (cf. Section \ref{sec:coord})
\bea
ds^2_2 &=& a^2 \left( r^2  dt^2 + \frac{dr^2}{r^2} \right) \;,  \nonumber\\
ds_2^2 &=& a^2 \left(  \sinh^2 \rho \, d \varphi^2 + d\rho^2 \right) \;\;\;,\;\;\; \varphi \equiv \varphi + 2 \pi
\;, \nonumber\\
ds_2^2 &=& \frac{a^2}{\sin^2 \sigma} \left( d \tau^2 + d \sigma^2 \right) \;\;\;,\;\;\;  - \pi  < \sigma < 0\;. 
\label{glline}
\eea
The first two line elements are related by the coordinate transformation \cite{Sen:2008yk},
\bea
\omega = \frac{1 - \frac{1}{r}  + i \, t}{1 + \frac{1}{r}  - i \, t } = e^{i \varphi} \,
\tanh \frac{\rho}{2} \;, 
\label{omph}
\eea
where here the coordinates $t, r, \rho$ are all dimensionless. Reinstating the dimensions of $t,r$ by
$t \rightarrow t/a $ and $r \rightarrow r/ a$, $\omega$ becomes
\bea
\omega = \frac{a - \frac{a^2}{r}  + i \, t}{a + \frac{a^2}{r}  - i \, t } \;.
\label{om}
\eea
When $r \rightarrow + \infty$, which implies $\rho \rightarrow + \infty$, this gives
\bea
t = a \, \tan \frac{\varphi}{2} \;.
\label{ttanv}
\eea
The second and the third line elements are related by \cite{Sen:2011cn}
\bea
\tan \left( \frac{\sigma + i \tau}{2} \right) = \tanh \left( \frac12 \left( \log \hat{\rho} + i \varphi \right) \right) \;\;\;,\;\;\; - \infty < \tau < + \infty \;\;\;,\;\;\; - \pi  < \sigma < 0 \;,
\label{stauhrv}
\eea
where
\bea
\hat{\rho} = \tanh \frac{\rho}{2} \;.
\eea
Hence, $\rho \rightarrow + \infty$ corresponds to $\hat{\rho} \rightarrow 1$. The segment 
$\hat{\rho} = 1, - \frac{\pi}{2} < \varphi < \frac{\pi}{2} $ gets mapped to the boundary $\sigma = 0$, while the 
segment 
$\hat{\rho} = 1,  \frac{\pi}{2} < \varphi < \frac{3\pi}{2} $ gets mapped to the boundary $\sigma = - \pi$ \cite{Sen:2011cn}. 

Combining \eqref{om} with \eqref{stauhrv} gives 
\bea
\tan \left( \frac{\sigma + i \tau}{2} \right) = \tanh \left( \frac12 \log \left( \frac{a - \frac{a^2}{r}  + i \, t}{a + \frac{a^2}{r}  - i \, t } \right) \right) \;,
\label{reltrtasi}
\eea
which relates Poincar\'e patch coordinates $(t,r)$ with the global coordinates $(\sigma, \tau)$. When $r = +\infty$ and $\sigma =0$, this becomes
\bea
\tan \left( i \frac{\tau}{2} \right) = i \tanh \left(\frac{\tau}{2} \right) = \tanh \left( i \frac{\varphi}{2} \right)
= i \tan \left( \frac{\varphi}{2} \right) = 
i \frac{t}{a} \;,
\label{reltvptau}
\eea
and hence
\bea
\frac{t}{a} = \tanh \left(\frac{\tau}{2} \right) \;.
\label{ttt}
\eea
On the other hand, when $r = +\infty$ and $\sigma = - \pi$, we obtain
\bea
\frac{t}{a} = \coth \left(\frac{\tau}{2} \right) \;.
\label{tct}
\eea

At large $r$ (and hence $x_0 \ll 1)$, we infer from \eqref{xrc}  (after rescaling $r \rightarrow r/a$) that $r = a/x_0$, while from \eqref{eax0} we obtain $\epsilon = a \, x_0$. By combining these two expressions, 
we infer that in Poincar\'e patch coordinates $1/\epsilon$ translates into the 
boundary location
\bea
 \frac{r}{a^2} = \frac{1}{\epsilon}  > 0 \;.
 \eea
Then, using \eqref{om}, we define
\bea
\omega_{\epsilon} (t) = \frac{a - \epsilon  + i \, t}{a + \epsilon  - i \, t } \;,
\eea
which satisfies $|\omega_{\epsilon} (t) |< 1$ so long as $\epsilon >0$. We may write $\omega_{\epsilon} (t)$ as
\bea
\omega_{\epsilon} (t) = \frac{a  + i \, \hat{t}}{a  - i \, \hat{t} } \;\;\;,\;\;
\hat{t} = t + i \epsilon \;.
\label{explom}
\eea
It follows that various  relations that were derived in \cite{Chamon:2011xk,Jackiw:2012ur} continue to hold by simply replacing $t$ by $\hat{t}$. In particular, we have
\bea
H  \,|t\rangle &=& -i \frac{d}{d t} 
|t\rangle \;, \nonumber\\
R  \,|t\rangle &=& -i \left( \frac{\hat{t}}{a} + \frac{a^2 + \hat{t}^2}{2a} \frac{d}{d t} \right)
|t\rangle \;, \nonumber\\
\mathbb{S}  \,|t\rangle &=& i \left( \frac{\hat{t}}{a} - \frac{a^2 - \hat{t}^2}{2a} \frac{d}{d t} \right)
|t\rangle \;.
\label{hrsran}
\eea
These can be verified explicitly using the relations given below, and we do so in Appendix \ref{sec:HRSt0}.  The state $|t\rangle$, constructed in \cite{Chamon:2011xk}, is defined as follows.
Following \cite{Chamon:2011xk} we introduce the operator
\bea
O(t)  = N(t)  \,
e^{ - \omega_{\epsilon}(t) \, L_+} \;\;\;,\;\;\; N(t) = 
\left( \frac{ \omega_{\epsilon} (t)+1}{2} \right)^2 \:.
\eea
Then, the state $|t \rangle$ is obtained by acting with the operator $O(t)$ on the $R$-vacuum $|0\rangle$,
\bea
|t\rangle = O(t) \, |0\rangle \;.
\label{opN}
\eea


As mentioned above, we identify the conformal quantum mechanics model living on each of the two boundaries of global Euclidean $AdS_2$ with the DFF model at coupling $g= \frac34$. We now construct states $|\tau \rangle$ living on either of the boundaries $\sigma = - \pi,0$.
We begin by considering the boundary $\sigma = 0$. In analogy with \eqref{ttt}, we define
\bea
\hat{t} = t + i \epsilon = 
a \, \tanh \frac{\hat{\tau}}{2}  \;\;\;,\;\;\;  \hat{\tau} = \tau + i \alpha \;.
\label{tvtauga}
\eea
Then, the right hand side of the expression for $\mathbb{S} |t\rangle$ in \eqref{hrsran} can be written as
\bea
- i \left( \cosh^2 \frac{\hat{\tau}}{2} \right)  \, 
\frac{d}{d \tau} \left( \frac{1}{\cosh^2 \frac{\hat{\tau}}{2}} \, |t \rangle \right) \;,
\eea
and hence the state 
\bea
|\tau \rangle \equiv \frac{1}{\cosh^2 \frac{\hat{\tau}}{2}} \, |t \rangle 
\label{taust}
\eea
satisfies \eqref{schrHRS}. This is a state at the boundary $\sigma = 0$.

On the boundary  $\sigma = - \pi$, we define in analogy with \eqref{tct},
\bea
\hat{t} = t + i \epsilon = 
a \, \coth \frac{\hat{\tau}}{2}  \;\;\;,\;\;\;  \hat{\tau} = \tau + i \alpha \;.
\label{tvtauga2}
\eea
Then, the right hand side of the expression for $\mathbb{S} |t\rangle$ in \eqref{hrsran} can be written as
\bea
- i \left( \sinh^2 \frac{\hat{\tau}}{2} \right)  \, 
\frac{d}{d \tau} \left( \frac{1}{\sinh^2 \frac{\hat{\tau}}{2}} \, |t \rangle \right) \;,
\eea
and hence the state 
\bea
|\tau \rangle \equiv \frac{1}{\sinh^2 \frac{\hat{\tau}}{2}} \, |t \rangle 
\label{taust2}
\eea
satisfies \eqref{schrHRS}. This is a state at the boundary $\sigma = - \pi$.

Note that these two states, which live at different boundaries, are distinct, except when $\tau \rightarrow \pm \infty$.

\subsection{The boundary state $|\tau \rightarrow - \infty\rangle$  }

We will now proceed to define boundary states $| \tau \rangle$ at $\tau \rightarrow - \infty$. Since the resulting expressions are divergent, we will need to regularise them.

First, we focus on the state $|\tau \rangle$ at the stretched  boundary $\sigma = - \delta$  
with $0 < \delta \ll 1$. We begin by determining $\alpha$ in \eqref{tvtauga}
and then take the limit $\tau \rightarrow - \infty$ on this boundary.
We work to first order in $\epsilon$.

Using \eqref{reltrtasi}, we infer
\bea
\tanh \left( \frac{- i \sigma +  \tau}{2} \right) = \tan \left( - \frac{i}{2} \log \omega_{\epsilon} (t) \right) \;\;\;,\;\;\;
\omega_{\epsilon} (t) =  \frac{1  + i \, \hat{t}/a}{
1 - i \, \hat{t} /a }  \;.
\eea
Expanding $\omega_{\epsilon} (t)$ to first order in $\epsilon$ gives
\bea
\log \omega_{\epsilon} (t) = \log \omega_0 (t) - \frac{2 a \epsilon}{a^2 + t^2} + {\cal O} (\epsilon^2) \;\;\;,\;\;\; 
\omega_0 (t) =  \frac{a  + i \, t}{
a - i \, t  } \;\;\;,\;\;\; | \omega_0 (t) | = 1 \;.
\eea
Using \eqref{omph}, we identify 
\bea
- \frac{i}{2} \log \omega_0 (t) = \frac{\varphi}{2} \;.
\eea
Then, using \eqref{ttanv} we
obtain, to first order in $\epsilon$,
\bea
\tanh \left( \frac{- i \sigma +  \tau}{2} \right) = \tan \left( \frac{\varphi}{2} 
+ i \frac{ a \epsilon}{a^2 + t^2} \right) = \frac{ \frac{t}{a} + i \tanh \left( \frac{ a \epsilon}{a^2 + t^2} \right) }{1 - i\,  \frac{t}{a} 
\, \tanh \left( \frac{ a \epsilon}{a^2 + t^2} \right) } = \frac{\hat{t} }{a} + {\cal O}(\epsilon^2) \,.
\label{ttanh}
\eea
Setting $\sigma = - \delta$ and comparing with \eqref{tvtauga}, we infer that
\bea
\alpha = \delta \;.
\eea
Next, we write the left hand side of \eqref{ttanh} as 
\bea
\tanh \left( \frac{- i \sigma +  \tau}{2} \right) = 1 - \frac{2}{e^{\tau} \, \cos \sigma + 1 - i e^{\tau} \, \sin \sigma  } = 1 - \frac{2 \left( e^{\tau} \cos \sigma + 1 + i e^{\tau} \sin \sigma \right) }{
\left( e^{\tau} \cos \sigma + 1\right)^2  + e^{2\tau} \sin^2 \sigma } \;, \nonumber\\
\eea
from which we infer the relations
\bea
\frac{t}{a} &=& 1 - \frac{2 \left( e^{\tau} \cos \sigma + 1 \right) }{
\left( e^{\tau} \cos \sigma + 1\right)^2  + e^{2\tau} \sin^2 \sigma } \;, \nonumber\\
\frac{\epsilon}{a} &=& -  2 \frac{e^{\tau} \sin \sigma  }{
\left( e^{\tau} \cos \sigma + 1\right)^2  + e^{2\tau} \sin^2 \sigma } \;.
\eea
At order $\delta$ and $e^{\tau} = e^{-1/T}$ with  $0 < \delta \ll 1$ and $0 < T \ll 1$, we obtain
 \bea
\frac{t}{a} &=& 
-1 + 2  e^{-1/T} \;, \nonumber\\
\frac{\epsilon}{a} &=& 
2  e^{-1/T} \delta > 0  \;.
\label{tepsstau}
\eea

Next, let us consider the state $|\tau \rangle$ at the stretched boundary $\sigma = - \pi + \delta$ with $0 < \delta \ll 1$. Then, setting  $\sigma = - \pi + \delta$ in \eqref{ttanh} and comparing with \eqref{tvtauga2}, we infer that
\bea
\alpha = -\delta \;.
\eea
Now we take the limit $\tau \rightarrow - \infty$.
At order $\delta$ and $e^{\tau} = e^{-1/T}$ with  $0 < \delta \ll 1$ and $0 < T \ll 1$, we obtain
 \bea
\frac{t}{a} &=& 
-1 - 2  e^{-1/T} \;, \nonumber\\
\frac{\epsilon}{a} &=& 
2  e^{-1/T} \delta > 0  \;.
\label{tepsstau2}
\eea

In what follows, we will set $t/a = -1$, while retaining a first order dependence on $\epsilon$ in all expressions. In the limit $\tau \rightarrow - \infty$, the two regularised boundary states that we have constructed are identical. We will 
denote this state by $|\delta \rangle$ and discuss its thermofield description next.

\subsection{Thermofield description of the state $|\delta \rangle$}

In order to write down the state $|\delta \rangle$ as a thermofield double state,
we will first
review the thermofield representation of the algebra \eqref{daffalg} by following \cite{10.1063/1.529540,Arzano:2023pnf}.

The algebra \eqref{daffalg} can be represented in terms of operators acting on the direct product of two independent
Hilbert spaces, $\mathcal{H}_L \otimes \mathcal{H}_R$, each described by a complete set of orthonormal states, which we denote by $S_L$ and $S_R$ respectively,
\bea
S_L = \{ |n\rangle_L , \, n \in \mathbb{N}_0 \} \;\;\;,\;\;\; S_R = \{ |n\rangle_R , \, n \in \mathbb{N}_0 \} \;,
\eea
with the states $| n\rangle_L$ and $| n\rangle_R$ satisfying
\bea
_{L}\langle m | n \rangle_L = \delta_{m,n} \;\;\;,\;\;\; _{R}\langle m | n \rangle_R = \delta_{m,n} \;.
\eea
The space $ \mathcal{H}_L \otimes \mathcal{H}_R$ is spanned by vectors belonging to the set
\bea
\{ |n,m\rangle = |n\rangle_L \otimes |m\rangle_R, \, \, n,m \in \mathbb{N}_0 \} 
\eea
satisfying
\bea
\langle n, m| l, k \rangle = \delta_{m,l} \, \delta_{n,k} \;.
\eea
We introduce two sets of annihilation and creation operators,
$a^{\dagger}_L, a^{\dagger}_R$ and $a_L, a_R$, respectively, satisfying
\bea
[a_{L}, a^{\dagger}_{L}] = 1 \;\;\;,\;\;\; [a_{R}, a^{\dagger}_{R}] = 1 \;,
\eea
with all other commutators vanishing. They act as follows on the states $|n,m\rangle$,
\bea
a_L |n,m\rangle = \sqrt{n} \, |n -1,m\rangle \;\;\;\,\;\;\; a^{\dagger}_{L} |n,m\rangle = \sqrt{n+1} \, |n +1,m\rangle
\;, \nonumber\\
a_R |n,m\rangle = \sqrt{m} \, |n,m -1\rangle \;\;\;\,\;\;\; a^{\dagger}_{R} |n,m\rangle = \sqrt{m+1} \, |n,m+1\rangle
\;,
\eea
and hence
\bea
a_L |0,m\rangle = a_R |n,0\rangle =  0 
\eea
as well as 
\bea
|n,m\rangle = \frac{\left( a^{\dagger}_L \right)^n}{\sqrt{n!}} \, \frac{\left( a^{\dagger}_R \right)^m}{\sqrt{m!}} \, 
|0\rangle_L  \otimes |0\rangle_R \;.
\eea
The generators of the algebra \eqref{daffalg} can be represented by
\bea
R = \frac12 \left( a^{\dagger}_L \, a_L + a^{\dagger}_R \, a_R + 1 \right) \;\;\;,\;\;\;\
L_+ =  a^{\dagger}_L \,  a^{\dagger}_R \;\;\;,\;\;\; L_- =  a_L \,  a_R \;.
\label{LpLmR}
\eea
A state $|n,m\rangle$ is an eigenstate of $R$ with eigenvalue $\frac12 (n + m + 1)$.

To obtain the thermofield double description of a state $|\tau\rangle$, we follow 
\cite{Arzano:2023pnf} and map the 
$R$-vacuum $|0\rangle$ in \eqref{Rgen}, which is an eigenstate of $R$ with eigenvalue $r_0=1$, to the groundstate of the operator $R$ given in \eqref{LpLmR}. The latter state, 
\bea
|0,0\rangle = |0\rangle_L \otimes |0\rangle_R \;,
\eea
is an eigenstate of $R$ in \eqref{LpLmR} with eigenvalue $1/2$. Similarly, the eigenstate $|n\rangle$ 
in \eqref{Rgen} with eigenvalue $n+1$ is mapped to the eigenstate $|n,n\rangle$ of weight $n + \frac12$, as shown above. We will see later that this is a consistent construction of a thermofield double description, as the entanglement entropy obtained by tracing out states in one of the Hilbert space copies is equal to the Boltzmann entropy computed in the original CQM.

Let us first consider 
the state $|\tau\rangle$ given in \eqref{taust}, whose thermofield double representation is
\bea
|\tau \rangle =  \frac{N(t)}{\cosh^2 \frac{\hat{\tau}}{2}} \, e^{ - \omega_{\epsilon}(t) \, L_+} \,
|0\rangle_L \otimes |0\rangle_R \;,
\label{taustR}
\eea
with $L_+$ given in \eqref{LpLmR}. We obtain
\bea
|\tau \rangle &=&  \frac{N(t)}{\cosh^2 \frac{\hat{\tau}}{2}} \, \sum_{n=0}^{+ \infty} \,
(-1)^n \, \omega^n_{\epsilon}(t) \, |n\rangle_L \otimes |n\rangle_R \nonumber\\
&=& \frac{a^2}{(a - i \hat{t} \, )^2 } \left( 1 - \frac{\hat{t}^2}{a^2} \right)
\, \sum_{n=0}^{+ \infty} \, 
(-1)^n \, \left( \frac{a  + i \, \hat{t}}{a  - i \, \hat{t} } \right)^n \, |n\rangle_L \otimes |n\rangle_R \;,
\label{taustR2}
\eea
where we used \eqref{tvtauga} to express
\bea
\frac{1}{\cosh^2 \frac{\hat{\tau}}{2}} = 1 - \frac{\hat{t}^2}{a^2} \;.
\eea
It follows that
\bea
\langle \tau |\tau \rangle &=& \frac{a^4}{| a - i \hat{t} \, |^4 }\, \Big| 1 - \frac{\hat{t}^2}{a^2} \Big|^2 \,
 \sum_{n=0}^{+ \infty} \, \left(  \frac{(a - \epsilon)^2  + t^2 }{ (a + \epsilon)^2  + t^2 } \right)^n
 \nonumber\\
&=&  \frac{a^4}{(a + \epsilon)^2  + t^2 } \,  \Big| 1 - \frac{\hat{t}^2}{a^2} \Big|^2  \, \frac{1}{4 a \, \epsilon} \;.
\eea
We define
\bea
{\cal N}^2 (t) \equiv \frac{ 4 a \, \epsilon \, \left((a + \epsilon)^2  + t^2\right) }{ | a^2 - \hat{t}^2|^2} \;.
\eea

Now we consider a state $|\tau \rangle$ defined at
$\tau \rightarrow - \infty$, which is consistent with setting $t= -a$ from \eqref{tepsstau}. In this limit,  we get 
\bea
{\cal N}^2 \equiv
{\cal N}^2 (t \rightarrow -a ) = \frac{ 4 a \, \epsilon \, \left((a + \epsilon)^2  + a^2\right) }{ | a^2 - 
(a - i \epsilon)^2|^2} = \frac{ 4 a \, 
\left((a + \epsilon)^2  + a^2\right) }{ \epsilon \,  (4 a^2  + \epsilon^2) }  \;.
\eea
To order $1/\epsilon$ this reads
\bea
{\cal N}^2 = \frac{2 a}{\epsilon} + {\cal O}(\epsilon^0) \;.
\eea

We define the normalised state $| \delta \rangle$ by
\bea
| \delta \rangle \equiv {\cal N}  \,  |\tau \rightarrow  - \infty \rangle \;\;\;,\;\;\; \langle \delta |\delta \rangle = 1 \;,
\eea
where 
\bea
|\tau \rightarrow  - \infty \rangle = 
\frac{ 2 i \epsilon a }{(a + \epsilon + i a  )^2 } 
\, \sum_{n=0}^{+ \infty} \, 
(-1)^n \, \left( \frac{a  - \epsilon - i a }{a + \epsilon + i a  } \right)^n \, |n\rangle_L \otimes |n\rangle_R \;.
\label{taustR4}
\eea
To lowest order in $\epsilon$ we have
\bea
\frac{a  - \epsilon - i a }{a + \epsilon + i a  } = - i \left( 1 - \frac{\epsilon}{a} \right) + {\cal O}(\epsilon^2) \;, \nonumber\\
 {\cal N}\, \frac{ 2 i \epsilon a }{(a + \epsilon + i a  )^2 } 
= \sqrt{ \frac{ 2 \epsilon}{a} } +  {\cal O}(\epsilon) \;.
\eea
Using these expressions, we obtain the following leading expression for $|\delta \rangle$ in an $\epsilon$-expansion,
\bea
| \delta \rangle = \sqrt{ \frac{ 2 \epsilon}{a} } \, \sum_{n=0}^{+ \infty} e^{- \frac{\epsilon}{a}  n + i \frac{\pi}{2}  n} \, |n\rangle_L \otimes |n\rangle_R \;,
\eea
which we recast as
\bea
| \delta \rangle = \frac{1}{\sqrt{Z(\beta)} } \; \sum_{n=0}^{+ \infty} e^{- \frac{\beta}{2} \, E_n + i \alpha_n} \, |n\rangle_L \otimes |n\rangle_R 
\label{dtherm2}
\eea
with 
\bea
E_n = n \;\;\;\;,\;\;\; \beta = \frac{2 \epsilon}{a} \;\;\;,\;\;\; \alpha_n = i \frac{\pi}{2} \, n \;.
\label{betid}
\eea
Here,  $Z(\beta) =  \displaystyle \sum_{n=0}^{+ \infty} e^{- \beta \, E_n}$ which, to lowest in $\beta$,
satisfies $1/Z(\beta) = \beta$. Note that at this order in $\beta$, 
$1/Z(\beta)$ equals $1/Z_{\rm CQM}(\beta)$, where $Z_{\rm CQM}$ denotes the partition function of the DFF model given in \eqref{zcqmdff}.
Also note that the boundary time periodicity $\beta$ is distinct from the bulk time periodicity \eqref{freethermo}.
Thus, at this order of the $\epsilon$-expansion, the state $|\delta \rangle$ describes a thermofield double state 
that involves time-independent phases $\alpha_n$; this differs from the recent work
\cite{Banerjee:2024fmh}, where the phases $\alpha_n$ are taken to be time-dependent and appear in the context of wormholes.

We defined the above state $| \delta \rangle$ by considering a state $|\tau \rangle$ on the boundary $\sigma = - \delta$
with $0 < \delta \ll 1$ and taking the limit $\tau \rightarrow - \infty$.
Now let us define a state $| \delta \rangle$ by 
considering a state $| \tau \rangle$ on the boundary $\sigma = - \pi + \delta$ and then again taken the limit $\tau \rightarrow - \infty$. To this end, we start from the state $|\tau \rangle $ given in 
\eqref{taust2} and proceed as above. Using  \eqref{tvtauga2} to express
\bea
\frac{1}{\sinh^2 \frac{\hat{\tau}}{2}} = \frac{\hat{t}^2}{a^2} -1 \;,
\eea
we find that the resulting state $|\delta \rangle$ equals \eqref{dtherm2}. The state $|\delta \rangle$ is thus shared by both boundaries of global Euclidean $AdS_2$.

 We view the two Hilbert spaces $\mathcal{H}_L \otimes \mathcal{H}_R$ as being associated to the two boundaries of global $AdS_2$ located at $\sigma = -\pi, 0$. By tracing over 
$\mathcal{H}_L $ associated with the boundary $\sigma = - \pi$, we obtain an 
 associated von Neumann entropy which is given by
\bea
S_{\rm vN} = - \log \beta  = - \log \frac{2 \epsilon}{a} \;,
\eea
as we will show below.

Since the boundary time periodicity $\beta$ goes to zero as $\epsilon$
goes to zero (cf. \eqref{betid}), $\epsilon$ serves as a UV length scale cutoff on the boundary. 
The IR length scale cutoff is set to $a/x_0 = a^2/\epsilon$.
This is consistent 
with the expected relation between the IR-UV cutoffs in $AdS/CFT$.

\subsection{Von Neumann entropy}

Having defined the state $|\delta \rangle$, we consider the pure state density matrix 
$|\delta\rangle \langle \delta |$ on $\mathcal{H}_L \otimes \mathcal{H}_R$, which we denote by 
$\hat{\rho}$,
\bea
\hat{\rho} &=& |\delta\rangle \langle \delta | = \frac{1}{Z(\beta) } 
\, \sum_{n,m=0}^{+ \infty} \, e^{- \frac{\beta}{2} \, \left( E_n + E_m \right)  + i \left( \alpha_n -
\alpha_m \right)} \, 
|n\rangle_L \langle m|_L \otimes |n\rangle_R \langle m|_R\;.
\eea
By tracing over $\mathcal{H}_L$ we obtain the reduced density matrix $\rho$,
\bea 
\rho =  \text{Tr}_{\mathcal{H}_L} \hat{\rho} = \sum_{l=0}^{+ \infty}    {_{L}
\langle} l | \hat{\rho}| l \rangle_L = \frac{1}{Z(\beta) } 
\, \sum_{l = 0}^{+ \infty} \, e^{- \beta \,  E_l  } \, | l \rangle_R \langle l |_R 
\label{redd1bd}
\eea
The associated entanglement entropy $S_{\rm vN}$ is
\bea
S_{\rm vN} = - \displaystyle \text{Tr}_{ \mathcal{H}_R} \, \rho \, \log \rho \;,
\eea
which can be computed using
\bea
S_{\rm vN} = - \lim_{k \rightarrow 1} \frac{d}{d k} \text{Tr}_{\mathcal{H}_R} \, {\rho}^k \;.
\label{repvN}
\eea
To do so \cite{Witten:2024upt}, one first computes ${\rho}^k$ with $k \in \mathbb{N}$,
\bea
\text{Tr}_{\mathcal{H}_R} \, {\rho}^k = \frac{Z( k \beta)}{Z^k(\beta)} \;,
\eea
and then analytically continues
the result to a holomorphic function on the half-plane ${\rm Re} \, k \geq 1 $. As discussed in \cite{Witten:2024upt}, this analytic continuation exists and is unique.
Using that $Z(\beta) = a/(2 \epsilon) $ to leading order in $\epsilon$, we obtain
\bea
\frac{Z( k \beta)}{Z^k(\beta)} = \frac{a}{2 \epsilon k } \left( \frac{2 \epsilon}{a} \right)^k \;,
\eea
and hence, in the limit $\epsilon \rightarrow 0$, the von Neumann entropy $S_{\rm vN} (\rho)$ is 
\bea
S_{\rm vN}  = - \log  \frac{ 2 \epsilon}{a } \,.
\label{vnentcqm}
\eea

The thermofield doubling of the Hilbert space ensures that the entanglement entropy 
\eqref{vnentcqm}
between two identical Hilbert space copies on the two boundaries of global $AdS_2$ equals the von Neumann entropy of the CQM on one of the boundaries. The latter is obtained as follows. We start from the partition function $Z_{\rm CQM}(u)$, given in \eqref{zcqmdff}, of the conformal quantum mechanics model on one of the boundaries of global $AdS_2$. We use \eqref{betid} and set
$u =\beta =  2 {\bar \epsilon} = 2 \epsilon/a$, in which case
$Z_{\rm CQM}({\bar \epsilon}) = e^{- \beta} Z(\beta) = e^{- \beta}
\displaystyle \sum_{n=0}^{+\infty} e^{- \beta \, E_n} = {\rm Tr} e^{- \beta R} $. The associated density matrix is 
$\rho = e^{- 2 {\bar \epsilon}  R}/Z_{\rm CQM}({\bar \epsilon})$, where $R$ denotes the compact generator \eqref{Rgen} with $r_0=1$.
Denoting the eigenvalues of $\rho$
by $\lambda_n = e^{- 2 {\bar \epsilon} \, (n+1)}/Z_{\rm CQM} ({\bar \epsilon}) $, the von Neumann entropy is
\bea
S_{\rm vN} = - \sum_{n=0}^{+\infty} \lambda_n \, \log \lambda_n 
&=&
\log Z_{\rm CQM} ({\bar \epsilon}) +  \frac{2{\bar \epsilon}}{Z_{\rm CQM} ({\bar \epsilon})} \,\sum_{n=1}^{+\infty}
n \, e^{- 2 {\bar \epsilon} \, n} \nonumber\\
&=& \log Z_{\rm CQM}({\bar \epsilon}) - 
{\bar \epsilon}
\, \partial_{\bar \epsilon} \log Z_{\rm CQM}({\bar \epsilon}) \;.
\eea
In the limit ${\bar \epsilon} \rightarrow 0^+$, $Z_{\rm CQM}({\bar \epsilon})$ behaves as $Z_{\rm CQM}({\bar \epsilon})\sim 1/(2{\bar \epsilon})$ (cf. \eqref{zcqmdff}) and hence
\bea
S_{\rm vN} = \log \frac{a}{2 \epsilon} \;,
\eea
up to an $\bar \epsilon$-independent constant. This expression agrees with the von Neumann entropy of entanglement between the two DFF copies on the boundaries of global $AdS_2$ given in \eqref{vnentcqm}.

Comparing with \eqref{finalent} and \eqref{dSL}, we establish that, up to $\epsilon$-independent constants,
\bea
{\cal S}_{\rm total} =\, 4\Delta S\,= \log \frac{L}{2 \pi a}  = S_{\rm vN} \;.
\label{dSSnn}
\eea
The last equality in \eqref{dSSnn}
expresses $ S_{\rm vN}$, the von Neumann
entropy of entanglement between the two CQM copies on the boundaries of global $AdS_2$,  in terms of 
a geometrical quantity $L$ in the bulk. We view this as the $AdS_2/CFT_1$ version of the Ryu-Takayanagi conjecture \cite{Ryu:2006bv,Ryu:2006ef} in our setup. 
The equality between ${\cal S}_{\rm total}$ and $ S_{\rm vN}$ equates  bulk entanglement entropy across black hole horizons to boundary entanglement entropy. 
We  note here that the evaluation of both the von Neumann entropy \eqref{repvN} in the boundary theory as well the bulk entanglement entropy \eqref{finalent} (based on the result \eqref{entflat} \cite{Calabrese:2004eu}) has been done using the replica trick. Both these entropies arise at the 1-loop level, in contrast to previous tree-level bulk replica computations of black hole entropy \cite{Azeyanagi:2007bj}.
We further note that \eqref{dSSnn} remains valid when the field theory on $AdS_2$  consists of $n$ multiplets, each of which contains a massless scalar field and a massless Majorana fermion.
The central charge of this system is then $c = \frac32 n$, and the conformal quantum mechanics model on each of the two boundaries of global $AdS_2$ is identified with $n$ copies of the DFF model.

\section{Concluding remarks}

In this paper, we have shown that the regularised one-loop effective action for a CFT with central charge $c = {\bar c} = 1 + \frac12 = \frac32$ in Euclidean $AdS_2$ spacetime contains the contribution $\Delta S = \frac{c}{6} \log \frac{a}{\epsilon}$,
 where $a$ denotes the $AdS_2$ scale and $\epsilon$ denotes the UV regulator. In this computation, equality of the bosonic and fermionic degrees of freedom in the field content is essential to cancel $\frac{1}{\epsilon^2}$ divergences leaving only the logarithmic divergent term $\Delta S$. 
  We have shown that $4 \Delta S$ equals the total entanglement entropy across the black hole horizons embedded in global Lorentzian $AdS_2$  and is exactly equal to the entanglement entropy between two copies of a $CFT_1$, each living on one of the two boundaries of global Euclidean $AdS_2$. The dual $CFT_1$ is a CQM identified by writing the unregularised bulk one-loop effective action as an integral, whose integrand is the requisite dual $CFT_1$ partition function.
    The identified dual $CFT_1$ in this case is the conformal quantum mechanics model of de Alfaro-Fubini-Furlan (DFF) \cite{deAlfaro:1976vlx} at the specific coupling value $g = 3/4$.  Further,
  the boundary entanglement entropy is expressed in terms of the logarithm of the length of a closed geodesic on the Euclidean $AdS_2$ disc. We view this relation between a measure of boundary quantum entanglement and a bulk geometrical property as the Ryu-Takayanagi conjecture in $AdS_2/CFT_1$ in our setup.
 \par Simultaneously, the identification of $4 \Delta S$ with bulk and boundary entanglement entropies can be viewed as a bulk-boundary entanglement entropy correspondence which relates quantum entanglement in the bulk to that to quantum entanglement in the boundary.

Our third result hinges on the observation that given the non-negative von Neumann entropy of the DFF model expressed as a function of a length scale $a$ and a short distance cutoff $\epsilon$, we can write the infinitesimal change in the entropy $S_{\rm vN}$ as 
\bea dS_{\rm vN}= - e^{S_{\rm vN}}\,  \frac{d\epsilon}{a}. \eea 
One can construct a $1+1$ space-time that encodes the above relation in its causal structure and is described by the metric 
\bea ds^2 = - e^{2 S_{\rm vN}} \, \frac{d \epsilon^2}{a
^2} + dS^2_{\rm vN} \;,
\eea  
which encodes a global  Lorentzian $AdS_2$ metric near one of its boundaries. The above metric can be thought of as the Fisher information metric on the information space of the von Neumann entropy and the UV scale. This observation is comparable to a related one in the context of a Calogero model as in \cite{Lechtenfeld:2015wka}.

One may wonder whether the DFF model also plays a role in providing a $CFT_1$ description of the logarithmic area corrections (if they exist) to the entropy of BPS black holes in four dimensions \cite{Banerjee:2010qc, Banerjee:2011jp,Sen:2012kpz}.
We emphasise here that black hole entropy is encoded in the 4D near horizon geometry, $AdS_2 \times S^2$,  which is computed in the quantum entropy function formalism as a path integral of fields which generically have legs in both $AdS_2$ and $S^2$.  The DFF entanglement entropy can hence only feed into the $AdS_2$ contribution to the black hole entropy. Nevertheless, as a future line of investigation, 2D entanglement entropy computations in JT gravity can potentially shed light on 2D wormholes that have been known 
to contribute to $R^2$ corrected black hole entropy in supersymmetric theories such as $N=4$ \cite{LopesCardoso:2022hvc}.

\section*{Acknowledgements}
We are grateful to Robert de Mello Koch for reading our manuscript and feedback.
We would like to thank Vishnu Jejjala, Amihay Hanany and Sameer Murthy for helpful discussions. 
The authors are supported
by FCT/Portugal 
through project UIDB/04459/2020 with DOI identifier 10-54499/UIDP/04459/2020.
The authors would like to thank the Isaac Newton Institute for Mathematical Sciences for support and hospitality during the program {\it Black holes: bridges between number theory and holographic quantum information} when work on this paper was undertaken; this work was supported by EPSRC grant number EP/R014604/1.

\appendix

\section{$ W_{B,F} (u)$ \label{sec:rewreff}}

Let us first consider the spectral density \eqref{densbos}. $\rho_B (\nu) $ has simple poles located at
\bea
\nu_n = i \left( n + \frac12 \right) \;\;\;,\;\;\;
n \in \mathbb{Z} \; .
\eea
Expanding $\nu$ around $\nu_n$, i.e. $
\nu = \nu_n + \delta \nu$, 
we infer that in the vicinity of the simple pole $\nu_n$,
\bea
\rho_B (\nu) = \frac{ \nu_n}{\pi \, \delta \nu} \:.
\eea
We deform the integration contour in \eqref{wu} to run along the two sides of the positive imaginary axis. We therefore restrict our attention to the poles $\nu_n$
with $n \in \mathbb{N}_0$.

The resulting contour integral in \eqref{wu} is then evaluated by summing over residues. Each simple pole contributes
\bea
2 \pi i \, e^{i \nu_n u}  \; \frac{ \nu_n}{\pi} = 
- (2 n + 1) \, e^{-(\frac12 + n) u} \;\;\;,\;\;\; n \in \mathbb{N}_0 \;.
\eea
Summing over all the residues gives (here we take ${\rm Re} \, u > 0$)
\bea
\sum_{n=0}^{\infty} (2n + 1) \, e^{ - n u} = - 2 \frac{d}{du} \frac{1}{1-e^{-u}} + \frac{1}{1-e^{-u}} = 2 \frac{e^{-u}}{(1-e^{-u})^2}  + \frac{1}{1-e^{-u}} = \frac{1 + e^{-u}}{(1-e^{-u})^2} \;, \nonumber\\
\eea
and hence we get \cite{Sun:2020ame}
\bea
W_{B}(u)= - e^{- \frac12 u}  \,
\frac{1 + e^{-u}}{(1-e^{-u})^2}  \;.
\eea

Next, we consider the spectral density \eqref{densrhof}.
$\rho_F (\nu)$  has simple poles located at
\bea
\nu_n = i n  \;\;\;,\;\;\; 
n \in \mathbb{Z} \;.
\eea
Expanding $\nu$ around $\nu_n$, i.e. $\nu = \nu_n + \delta \nu$, 
we infer that in the vicinity of the simple pole $\nu_n$,
\bea
\rho_F(\nu) = 2 \frac{ \nu_n}{\pi \, \delta \nu} \:.
\eea
We deform the integration contour in \eqref{wu} to run along the two sides of the positive imaginary axis. We therefore restrict our attention to the poles $\nu_n$
with $n \in \mathbb{N}$.

The resulting contour integral in \eqref{wu} is again evaluated by summing over residues. Each simple pole contributes
\bea
4 \pi i \, e^{i \lambda_n u}  \; \frac{ \lambda_n}{\pi} = 
- 4 n  \, e^{-n u} \;\;\;,\;\;\; n \in \mathbb{N} \;.
\eea
Summing over all the residues gives (here we take ${\rm Re} \, u > 0$)
\bea
\sum_{n=1}^{\infty}  4 n  \, e^{ - n u} = - 4 \frac{d}{du} \frac{1}{1-e^{-u}}  = 4 \frac{e^{-u}}{(1-e^{-u})^2} \;,
\eea
and hence we get \cite{Sun:2020ame}
\bea
W_{F}(u) =  - 4   \,
\frac{e^{-u}}{(1-e^{-u})^2}  \;.
\eea

\section{Regularising the one-loop effective action \label{sec:regonel}}

The one-loop effective action $\Gamma_{\rm 1-loop}$ for a field can be expressed in terms of the heat kernel $K(s)$ for the field,
\bea
\Gamma_{\rm 1-loop} 
= - \frac12 \int_0^{+\infty} \, \frac{ds}{s} \, K(s) \;.
\label{gam1Kcont}
\eea
This is a formal expression: it suffers from divergences, and hence needs to be regularised. It possesses two types of divergences: an infrared (IR) divergence associated with $s \rightarrow + \infty$, and an ultraviolet (UV) divergence associated with $s \rightarrow 0$. A standard regularization \cite{Vassilevich:2003xt} of the UV divergence consists in introducing a regulator $\epsilon >0$
and replacing \eqref{gam1Kcont} by
\bea
\Gamma^{\epsilon}_{\rm 1-loop} = - \frac12 \int_{\epsilon}^{+\infty} \, \frac{ds}{s} \, K(s) \;,
\label{gam1Kreg}
\eea
which then still requires the introduction of a further regulator to deal with the divergence associated with $s \rightarrow + \infty$.
Rather then introducing the regulator $\epsilon$ as in 
\eqref{gam1Kreg}, we will follow \cite{Anninos:2020hfj,Sun:2020ame} and introduce it in the form
\bea
\Gamma^{\epsilon}_{\rm 1-loop} = - \frac12 \int_{0}^{+\infty} \, \frac{ds}{s} \, e^{ - \epsilon^2/4s} \, K(s) \;.
\label{gam1Kreg2}
\eea
This will then be further supplemented with an infrared regulator.

Consider the expression
\bea
K(s) = \int_0^{+\infty} d \nu \, \rho(\nu) \, e^{ -  (\lambda^2 + \nu^2) s } \,,
\label{Keven}
\eea
with $\rho$ an even function, $\rho(-\nu) = \rho(\nu)$,  and $\lambda \geq 0$. Comparing with \eqref{Kb}  and \eqref{Kf}, we have dropped the normalisation factor $\pm V/(2 \pi a^2)$ and we have replaced $\bar s$ and $\bar \epsilon$ by $s$ and $\epsilon$ for notational simplicity. These factors can be easily reinstated in the final expression 
\eqref{gamWu5} given below.

Since the integrand in 
\eqref{Keven} is an even function, we first write it as
\bea
K(s) =\frac12 \,  \int_{-\infty}^{+\infty} d \nu \, \rho(\nu) \, e^{ -  (\lambda^2 + \nu^2) s } 
\label{Kinfinf}
\eea
and proceed to rewrite it as follows \cite{Sun:2020ame}.
Using the Gaussian integral
\bea
\int_{-\infty}^{+\infty} du \, e^{- \frac{u^2}{4s} + i \nu u } = 
\sqrt{4 \pi s} \, e^{-   \nu^2 s  } \;\;\;,\;\;\; 
s \in \mathbb{R}^+ \;,
\eea
we write \eqref{Kinfinf} as 
\bea
K(s) = \frac{1}{4 \sqrt{\pi \, s} } \,  \, e^{ -   \lambda^2 s } \, \int_{-\infty}^{+\infty} d \nu \, \left[\rho(\nu) 
 \int_{-\infty}^{+\infty} du \,  e^{- \frac{u^2}{4s} + i \nu u } 
 \right] \;.
 \eea
Next, assuming that we can interchange the two integrations, we obtain
\bea
K(s) = \frac{1}{4 \sqrt{\pi \, s} } \, e^{ -   \lambda^2 s } \, \int_{-\infty}^{+\infty} du \,
 e^{- \frac{u^2}{4s} } \, W(u) \,,
 \label{KW}
 \eea
where $W$ denotes the Fourier integral of $\rho$,
\bea
W(u) \equiv \int_{-\infty}^{+ \infty}
d\nu \, \rho(\nu ) \, e^{i \nu u } \;.
\label{wu2}
\eea
Since $\rho (- \nu) = \rho (\nu)$, we formally have $W(-u) = W(u)$.
Then, inserting \eqref{KW} into \eqref{gam1Kreg2} and interchanging the
integrations over $s$ and $u$ we obtain
\bea
\Gamma^{\epsilon}_{\rm 1-loop} = - \frac{1}{8 \sqrt{\pi } }
\, \int_{-\infty}^{+ \infty} du \, W(u) \, \left[
\int_{0}^{\infty} \, \frac{ds}{s^{3/2}} \, e^{ - \frac{(\epsilon^2 + u^2)}{4s} - \lambda^2 s } \right] \;.
\label{intWs}
\eea
The integral over $s$ is proportional to the modified Bessel function of the second kind of degree $1/2$ \cite{modbess},
\bea
K_n (z) &=& \frac12 \left( \frac12 z \right)^{n} \, \int_0^{+\infty} \, \frac{ds}{s^{n+1}} \, e^{ - \frac{z^2}{4s} - s } \;\;\;,\;\;\; |{\rm arg} \, z | < \frac{\pi}{4} \;, \nonumber\\
K_{1/2} (z) &=& \sqrt{\frac{\pi}{2}} \, \frac{e^{-z}}{\sqrt{z}} \;.
\eea
When $\lambda > 0$, we obtain, after performing the rescaling $ \lambda^2 s \rightarrow s$ in \eqref{intWs},
\bea
\Gamma^{\epsilon}_{\rm 1-loop} = - \frac{1}{4  }
\, \int_{-\infty}^{+\infty} \frac{du}{\sqrt{\epsilon^2 + u^2}} \, W(u) \, e^{- \lambda \sqrt{\epsilon^2 + u^2}} \;.
\label{gamWu}
\eea
When $\lambda = 0$, we define $\Gamma^{\epsilon}_{\rm 1-loop}$ by setting $\lambda =0$ in \eqref{gamWu}.

This expresses the one-loop effective action in terms of the Fourier integral $W$ of the spectral density $\rho$.
However, some of the assumptions that we made in order to arrive at \eqref{gamWu} are invalid. 
As a consequence, expression \eqref{gamWu} is ill-defined:
since the spectral density $\rho (\nu)$ in \eqref{densbos} and in \eqref{densrhof}
does not decay to zero as $\nu \rightarrow \pm \infty$, the Fourier integral \eqref{wu2} is divergent. To deal with this divergence, we proceed as follows \cite{Anninos:2020hfj,Sun:2020ame}.
Since, in the examples that we study, $\rho(\nu)$ only has simple poles located on the imaginary axis of the complex plane $\nu$ and none
on the real axis, we deform the integration contour in \eqref{wu2} to run along the two sides of the positive imaginary axis. 
The resulting contour integral is then evaluated by summing over residues.
The expression that one obtains for $W(u)$ in this manner has a pole at $u =0$, cf. \eqref{dpW}.
To deal with this pole, we 
change the integration contour in \eqref{gamWu} to run over $\mathbb{R} \pm i \delta$, where
$\delta >0$ is taken to be small,
\bea
\Gamma^{\epsilon}_{\rm 1-loop} = - \frac{1}{8  } 
\left(  \int_{\mathbb{R} + i \delta} + \int_{\mathbb{R} - i \delta} \right)
\frac{du}{\sqrt{\epsilon^2 + u^2}} \, W(u) \, e^{- \lambda \sqrt{\epsilon^2 + u^2}} \;.
\label{gamWu2}
\eea
Note that there are two branch cuts in the complex plane $u$, due to the presence in \eqref{gamWu2} of the factor $\sqrt{ \epsilon^2 + u^2}$. Following \cite{Sun:2020ame},
we orient the two branch cuts as follows. One of them starts at $i \epsilon$ and runs to $+ i \infty$
along the positive imaginary axis of the complex plane $u$, while the other starts at $-i \epsilon$ and runs
to $ - i \infty$ along the negative imaginary axis of the complex plane $u$, as shown in Figure 5.1a of \cite{Sun:2020ame}.
Therefore, the regulator $\delta$ has to be taken to be smaller than $\epsilon$, to ensure that the integration contour does not cross the branch cuts. Since the integrand in \eqref{gamWu2} is invariant under $u \rightarrow -u$ due to $W(-u) = W(u)$, the pole at $u=0$ does not contribute, i.e. the two integration contours are equivalent, and hence
we write
\bea
\Gamma^{\epsilon}_{\rm 1-loop} = - \frac{1}{4  } 
 \int_{\mathbb{R} + i \delta} 
\frac{du}{\sqrt{\epsilon^2 + u^2}} \, W(u) \, e^{- \lambda \sqrt{\epsilon^2 + u^2}} \;.
\label{gamWu3}
\eea

This expression still requires a regularization in the infrared, which corresponds to the behaviour at  $|u| \rightarrow + \infty$.
Following \cite{Sun:2020ame}, we therefore insert an infrared regulator $\kappa > 0$ into \eqref{gamWu2}, to obtain the regularised
one-loop partition function
\bea
\Gamma^{\rm reg}_{\rm 1-loop} = - \frac{1}{4  } 
 \int_{\mathbb{R} + i \delta} 
\frac{du}{\sqrt{\epsilon^2 + u^2}} \, W(u) \, e^{- \lambda \sqrt{\epsilon^2 + u^2}} \, e^{- \kappa |u|} \;.
\label{gamWu4}
\eea
Introducing the notation 
\bea
H_{\lambda r } (u) \equiv -  W(u) \, e^{-  \lambda r  u} \;\;\;,\;\;\; r = \frac{\sqrt{\epsilon^2 + u^2}}{u} \;,
\eea
we express $\Gamma^{\rm reg}_{\rm 1-loop}$ as
\bea
\Gamma^{\rm reg}_{\rm 1-loop} =  \frac{1}{4  } 
 \int_{\mathbb{R} + i \delta} 
\frac{du}{r u }
\, H_{\lambda r } (u) \, e^{- \kappa |u|} \;.
\label{gamWu5}
\eea

\section{Evaluating the regularised one-loop effective action \label{sec:evreg1}}

Here we evaluate the regularised one-loop effective action \eqref{regcomb} following
 \cite{Anninos:2020hfj,Sun:2020ame}.
We write it in the form \eqref{gamWu5}, 
where we reinstate the normalisation factor $V/(2 \pi a^2)$ and 
replace $\epsilon$ by $\bar \epsilon = \epsilon/a$,
\bea
\Gamma^{\rm reg}_{\rm 1-loop} ( {\bar \epsilon} ) =  \frac{V}{8 \pi a^2}
 \int_{\mathbb{R} + i \delta} 
\frac{du}{r u }
\, {\tilde H} (r, u) \, e^{- \kappa |u|} \;,
\eea
where
\bea
{\tilde H} (r, u) =  - W_B(u)  \, e^{- \frac12 r u  } + \frac12 W_F (u) = 
\frac{1 + e^{-u}}{(1-e^{-u})^2} \,  e^{- \frac12 u}  \, e^{- \frac12 r u  } 
- 2  \frac{e^{-u}}{(1-e^{-u})^2}      \:.
\label{Hueps}
\eea
We define $H(u)$ by 
setting $\epsilon = 0$ in \eqref{Hueps}, 
\bea
H(u) = - \frac{e^{-u}}{1-e^{-u}} \:,
\label{Hu}
\eea
which we expand in a Laurent series around $u=0$,
\bea
\frac12 H ( u)  =  \sum_{n=0}^{+\infty} b_n \, u^{n-2} \;.
\eea
We then 
split $H(u)$ into \cite{Sun:2020ame}
\bea
H(u) = H^{\rm UV} (u) + H^{\rm IR}(u) \;,
\eea
where
\bea
\frac12 H^{\rm UV} (u)  =  \sum_{n=0}^2 b_n  \, u^{n-2} = \frac{b_0}{u^2} + \frac{b_1}{u} + b_2 
\eea
with
\bea
b_0 = 0 \;\;\;,\;\;\; b_1   =  - \frac12 \;\;\;,\;\;\; b_2  = \frac14 \;.
\eea
Similarly, by expanding ${\tilde H} (r, u)$ in a Laurent series around $u=0$ keeping $r$ fixed, 
we define
\bea
\frac12 {\tilde H}^{\rm UV} (r, u) &=& \sum_{n=0}^2 a_n (r) \, u^{n-2} = \frac{a_0(r)}{u^2} + \frac{a_1(r)}{u} + a_2(r) \nonumber\\
&=& 
\frac{1}{u^2} \left( -  \frac12 \, r \, u + \frac18 \left( 1 +  r^2 \right) u^2 \right) 
\eea
and split ${\tilde H}(r,u)$ into 
\bea
{\tilde H}(r, u) = {\tilde H}^{\rm UV} (r, u) + {\tilde H}^{\rm IR}(r, u) \;.
\eea
Using this, we split $\Gamma^{\rm reg}_{\rm 1-loop}$ 
into a UV part and an IR part \cite{Sun:2020ame},
\bea
\Gamma^{\rm reg}_{\rm 1-loop} = \Gamma^{\rm reg, UV}_{\rm 1-loop} + \Gamma^{\rm reg, IR}_{\rm 1-loop} \;,
\eea
where
\bea
\Gamma^{\rm reg, UV}_{\rm 1-loop} &=& 
\frac{V}{8 \pi a^2}
 \int_{\mathbb{R} + i \delta} 
\frac{du}{r u }
\, {\tilde H}^{\rm UV} (r, u) \, e^{- \kappa |u|} \;, \nonumber\\
\Gamma^{\rm reg, IR}_{\rm 1-loop} &=& 
\frac{V}{8 \pi a^2}
 \int_{\mathbb{R} + i \delta} 
\frac{du}{r u }
\, {\tilde H}^{\rm IR} (r, u) \, e^{- \kappa |u|} \;.
\eea
The UV part refers to the part of the one-loop effective action that would diverge when switching off the UV regulator $\epsilon$. Note that the UV part may also be divergent in the IR.  
The IR part refers to the part of the one-loop effective action that would diverge
when switching off the IR regulator $\kappa$. 
Since the IR part is finite in the UV, we may remove the
UV regulators $\delta$ and $\epsilon$
in the expression for $\Gamma^{\rm reg, IR}_{\rm 1-loop}$ and obtain
\bea
\Gamma^{\rm reg, IR}_{\rm 1-loop} =
 \frac{V}{4 \pi a^2} \, \int_0^{+\infty} 
\frac{du}{u } \, H^{\rm IR} (u) \, e^{- \kappa u} \;.
\eea
The latter can be evaluated in terms of the character zeta function \cite{Anninos:2020hfj,Sun:2020ame}
\bea
\zeta (z) = \frac{1}{\Gamma(z)} \int_0^{+\infty} \frac{du}{u} \, u^z \, H (u) \;,
\label{charz}
\eea
to give (see eeq. (C.15) in \cite{Anninos:2020hfj})
\bea
\Gamma^{\rm reg, IR}_{\rm 1-loop} = \frac{V}{2 \pi a^2} 
\left( \frac12 \zeta' (0) + b_2 \log \kappa \right) \;.
\eea
Expanding \eqref{Hu} in powers of $e^{-u}$, with $u>0$, 
\bea
H(u) =  - \sum_{n=0}^{+\infty} e^{- (n +1) u} \;,
\eea
inserting this into \eqref{charz} and evaluating the integral \eqref{charz} for  ${\rm Re} \, z > 0$ gives 
\bea
\zeta (z) = -  \sum_{n=0}^{+\infty} \frac{1}{(n+1)^z} = - \zeta_H (z, 1) \;,
\eea
where $\zeta_H$ denotes the Hurwitz zeta function $\zeta_H (z, \Delta)$ with $\Delta = 1$.
Using 
\bea
\zeta'_H (0, \Delta) = \log \Gamma (\Delta) - \frac12 \log (2 \pi) \;,
\eea
we obtain 
\bea
\zeta' (0) = -\zeta'_H (0,  1) =  \frac12 \log (2 \pi) \;.
\eea

Next we evaluate $\Gamma^{\rm reg, UV}_{\rm 1-loop}$,
\bea
\frac12 \int_{\mathbb{R} + i \delta} \, \frac{du}{2 r u }
{\tilde H}^{\rm UV} (r, u) \, e^{- \kappa \, |u|} =
\frac12 \int_{\mathbb{R} + i \delta} \, du \left[ -  \frac{1}{ 2 u^2}  + \frac18 \left( \frac{1}{r \, u} +  \, \frac{r^2 }{r \, u}  \right) \right]
\, e^{- \kappa \, |u|} \;.
\eea
By closing the contour of integration in the upper half plane, the integral over the first term vanishes, and we are left with
\bea
\frac{1}{16} \int_{\mathbb{R} + i \delta} \, du    \left( \frac{1}{r \, u} +  \frac{ r^2}{r \, u}  \right) 
\, e^{- \kappa \, |u|} \;,
\eea
which we write as 
\bea
\frac{1}{16} \int_{\mathbb{R} + i \delta} \, du    \left( \frac{2}{\sqrt{u^2 +1}} + \frac{1 }{u^2 \sqrt{u^2 +1}}   \right) 
\, e^{- \kappa \, {\bar \epsilon}\, |u|} \;.
\label{int12}
\eea
The second integral in \eqref{int12} does not require the IR regulator $\kappa$, and we obtain
\bea
\frac{1}{16} \int_{\mathbb{R} + i \delta} \, du  
\, \frac{1 }{u^2 \sqrt{u^2 +1}}  = - \frac18 \;.
\eea
The first integral in \eqref{int12} requires the IR regulator $\kappa$. 
First we note that we may remove the regulator $\delta$, in which case this integral becomes
\bea
\frac14 \int_0^{+\infty} du \,  \frac{1}{\sqrt{u^2 +1}} 
\, e^{- \kappa \, {\bar \epsilon}\, u} \;.
\eea
In the limit $\kappa \rightarrow 0^+$, this evaluates to
\bea
 \frac14 \int_0^{+\infty}  \, du \, \frac{1}{\sqrt{u^2 +1}} 
\, e^{- \kappa \, {\bar \epsilon} \, u} = - \frac14 \log \left( \frac{e^{\gamma} \kappa \, {\bar \epsilon}}{2} \right) \;,
\eea
where $\gamma$ denotes the Euler-Mascheroni constant.

Collecting all the terms, we obtain for the regularised one-loop effective action \eqref{regcomb},
\bea
\Gamma^{\rm reg}_{\rm 1-loop} (\bar{\epsilon} ) = 
-\log Z^{\rm reg}_{\rm 1-loop} (\bar{\epsilon} ) = \frac{V}{2 \pi a^2} \left[ - b_2 \log \left( \frac{e^{\gamma} \, \bar{\epsilon}}{2} \right) 
- \frac12 \zeta'_H (0,  1) - \frac18 
\right] \;,\;\bar{\epsilon} = \frac{\epsilon}{a} \;,\; b_2 = \frac14  \;. \nonumber\\
\eea
Note that the IR regulator $\kappa$ has canceled out in the log-term \cite{Sun:2020ame}.
Similar calculations were done in \cite{Grewal:2021bsu}. 

\section{The action of $H, R, \mathbb{S}$ on $|t \rangle$ \label{sec:HRSt0}}

To verify the relations \eqref{hrsran}, we express $H, \mathbb{S}, R$ as
\bea
H &=& \frac{1}{2a} \left( 2 R - L_+ - L_- \right) \;, \nonumber\\
\mathbb{S} &=& - \frac12 \left( L_+ + L_- \right) \;, \nonumber\\
R &=& - \mathbb{S} + a \, H \;.
\eea
To simplify notation, let us denote $\omega_{\epsilon}(t)$ by $\omega$ in the following. 
Then,
\bea
H  \, e^{ - \omega L_+ } |0\rangle = \frac{1}{2a} \left( 2 R \,  e^{ - \omega L_+}   + \frac{d}{d \omega}
e^{ - \omega L_+}  - [L_-,  e^{ - \omega L_+ }]
\right) |0\rangle \;.
\eea
Now we use
\bea
 [L_-,  e^{ - \omega L_+ }] = e^{ - \omega L_+} \left( - 2 \omega R + \omega^2 L_+ \right) \;,
 \eea
to obtain, with $R |0\rangle = |0\rangle$,
\bea
H  \, e^{ - \omega L_+ } |0\rangle = 
\frac{1}{2a} \left( 2 R \,  e^{ - \omega L_+}   + \frac{d}{d \omega}
e^{ - \omega L_+}  - e^{ - \omega L_+ } \left( - 2 \omega  + \omega^2 L_+ \right) 
\right) |0\rangle \;.
\eea
Next, using
\bea
 R \,  e^{ - \omega L_+} |0\rangle =  [R ,  e^{ - \omega L_+}] |0\rangle +  e^{ - \omega L_+} |0\rangle \;,
 \eea
and
\bea
[R ,  e^{ - \omega L_+}] |0\rangle = - \omega L_+ \, e^{ - \omega L_+} |0\rangle = \omega \frac{d}{d \omega} e^{ - \omega L_+} |0\rangle \;,
\eea
we get
\bea
 R \,  e^{ - \omega L_+} |0\rangle =  \left( 1 + \omega \frac{d}{d \omega} \right) e^{ - \omega L_+} 
|0\rangle \;.
\eea
It follows that 
\bea
H  \, e^{ - \omega L_+ } |0\rangle = \frac{1}{2a} \left(  2 ( 1 + \omega) e^{ - \omega L_+} 
|0\rangle + (1 + \omega)^2 \frac{d}{d \omega} e^{ - \omega L_+} 
|0\rangle \right) \;.
\eea
Hence, using \eqref{opN} we obtain
\bea
H  \,|t\rangle &=& \frac{1}{2a} \left(  2 ( 1 + \omega) 
|t\rangle + N \, (1 + \omega)^2 \frac{d}{d \omega} \left( \frac{1}{N} \, 
|t\rangle \right)  \right) \nonumber\\
&=& \frac{1}{2a} \left(  2 ( 1 + \omega) 
|t\rangle +  (1 + \omega)^2 \frac{d}{d \omega} 
|t\rangle \right)  + (1 + \omega)^2 N  \frac{d  }{d \omega}  \left( \frac{1}{N} \right) 
|t\rangle \;,
\eea
and using
\bea
N  \frac{d  }{d \omega}  \left( \frac{1}{N} \right) = - \frac{2}{ \omega +1} \;,
\eea
we get
\bea
H  \,|t\rangle = \frac{1}{2a} \,  (1 + \omega)^2 \frac{d}{d \omega} 
|t\rangle \;.
\eea
Now we use the explicit form of $\omega$ given in \eqref{explom} to infer
\bea
\omega + 1 &=& \frac{2a}{a- i \hat{t}} \;, \nonumber\\
\frac{d \omega}{dt} &=& \frac{2 i \, a }{ (a - i \hat{t} )^2} \;,
\eea
and hence 
\bea
H  \,|t\rangle = -i \frac{d}{d t} 
|t\rangle \;.
\eea

Similarly, we express $\mathbb{S} \,|t\rangle$ as
\bea
\mathbb{S} \,|t\rangle &=& \frac{\omega -1}{\omega + 1} \, |t\rangle + \frac{(1 + \omega)^2}{2} \, \frac{d}{d \omega} 
|t\rangle \nonumber\\
&=&  i \left( \frac{\hat{t}}{a} - \frac{a^2 - \hat{t}^2}{2a} \frac{d}{d t} \right)
|t\rangle \;,
\label{schrSt}
\eea
and finally,
\bea
R \,|t\rangle = \left( - \mathbb{S} + a \, H \right) \,|t\rangle =  -i \left( \frac{\hat{t}}{a} + \frac{a^2 + \hat{t}^2}{2a} \frac{d}{d t} \right)
|t\rangle \:.
\eea


\providecommand{\href}[2]{#2}\begingroup\raggedright\endgroup

\end{document}